\documentclass[conference]{IEEEtran}
\IEEEoverridecommandlockouts

\usepackage[numbers]{natbib}
\usepackage{textcomp}
\usepackage{float}
\usepackage{booktabs}
\usepackage{amsmath,amssymb,amsfonts}
\usepackage{graphicx}
\usepackage{multirow}
\usepackage{needspace} 
\usepackage{booktabs}
\usepackage{algorithm}
\usepackage{algpseudocode}
\usepackage{diagbox}
\usepackage{tabularx}
\usepackage{nicematrix,enumitem,booktabs}
\usepackage{verbatim}
\usepackage[colorinlistoftodos]{todonotes}
\newtheorem{definition}{Definition}

\begin{document}

\newcommand{\tp}[1]{{\color{red} {\bf ??? #1 ???}}\normalcolor}

\title{Fast and Exact Similarity Search in less than a Blink of an Eye}

\author{
    \IEEEauthorblockN{Patrick Sch\"afer\IEEEauthorrefmark{1}, Jakob Brand\IEEEauthorrefmark{2}, Ulf Leser\IEEEauthorrefmark{3}}    
    \IEEEauthorblockA{\textit{Humboldt-Universit\"at zu Berlin} \\    
    Berlin, Germany \\
    Email: \IEEEauthorrefmark{1}patrick.schaefer@hu-berlin.de,\\
    \IEEEauthorrefmark{2}brandjak@hu-berlin.de,\\
    \IEEEauthorrefmark{3}leser@informatik.hu-berlin.de}
    \and
    \IEEEauthorblockN{Botao Peng}
    \IEEEauthorblockA{\textit{Institute of Computing Technology} \\
    \textit{Chinese Academy of Sciences} \\
    Beijing, China \\
    pengbotao@ict.ac.cn}
    \and
    \IEEEauthorblockN{Themis Palpanas}
    \IEEEauthorblockA{\textit{LIPADE} \\
    \textit{Universit\'e de Paris}\\
    Paris, France \\
    themis@mi.parisdescartes.fr}    
}

\maketitle

\begin{abstract}
Similarity search is a fundamental operation for analyzing data series (DS), which are ordered sequences of real values. To enhance efficiency, summarization techniques are employed that reduce the dimensionality of DS. SAX-based approaches are the state-of-the-art for exact similarity queries, but their performance degrades for high-frequency signals, such as noisy data, or for high-frequency DS.
In this work, we present the SymbOlic Fourier Approximation index (SOFA), which implements fast, exact similarity queries. SOFA is based on two building blocks: a tree index (inspired by MESSI) and the SFA symbolic summarization. It makes use of a learned summarization method called Symbolic Fourier Approximation (SFA), which is based on the Fourier transform and utilizes a data-adaptive quantization of the frequency domain. To better capture relevant information in high-frequency signals, SFA selects the Fourier coefficients by highest variance, resulting in a larger value range, thus larger quantization bins. The tree index solution employed by SOFA makes use of the GEMINI-approach to answer exact similarity search queries using lower bounding distance measures, and an efficient SIMD implementation.
We further propose a novel benchmark comprising $17$ diverse datasets, encompassing 1 billion DS. Our experimental results demonstrate that SOFA outperforms existing methods on exact similarity queries: it is up to 10 times faster than a parallel sequential scan, 3-4 times faster than FAISS, and 2 times faster on average than MESSI. For high-frequency datasets, we observe a remarkable 38-fold performance improvement. 
\end{abstract}

\begin{IEEEkeywords}
Data Series, Time Series, Similarity Search, Exact, Euclidean Distance.
\end{IEEEkeywords}

\sloppy

\section{Introduction}

The advancements and deployments of modern sensors have led to the generation, collection, and analysis of massive datasets of data series (DS) in nearly every scientific field~\cite{DBLP:journals/sigmod/Palpanas15,Palpanas2019,DBLP:journals/dagstuhl-reports/BagnallCPZ19,DBLP:conf/wims/EchihabiZP20}. 
A DS is an ordered sequence of real values. The most common type of DS is the time series, which is ordered in time. Other frequent orders are wave lengths, angles, pixel offsets, or locations~\cite{dau2019ucr}.

Important analysis problems for data series are querying~\cite{echihabi2020lernaean}, classification~\cite{middlehurst2024bake}, clustering~\cite{holder2024review}, anomaly detection~\cite{wenig2022timeeval}, which all require special algorithms that take the sequential nature of the values into account. At the heart of these techniques is similarity search, which aims to identify the series in a dataset that is closest to a given query series based on a distance measure, such as the widely used Euclidean distance (ED). Similarity search can be split into two main categories: exact search and approximate search~\cite{echihabi2020lernaean}. In this work, we focus on exact similarity search using ED.

\begin{figure*}[t]
    \centering  
	\includegraphics[width=2.00\columnwidth]{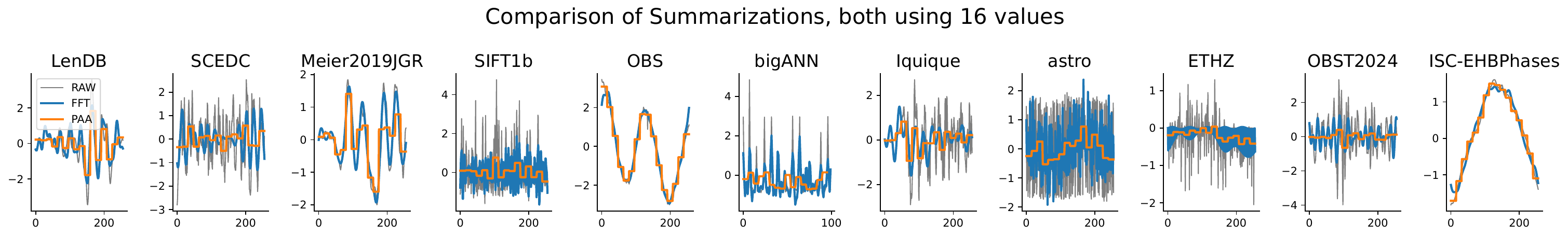}
    \includegraphics[width=2.00\columnwidth]{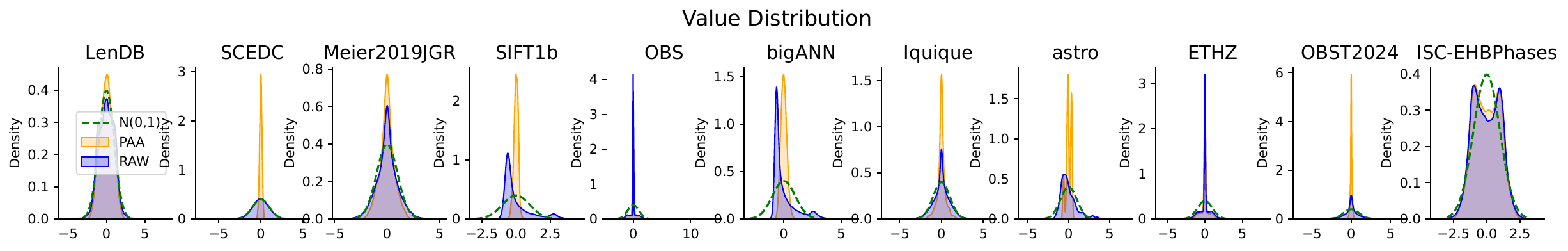}
	\caption{
	TOP: PAA (orange) fails to approximate a data series (in gray) with high frequency, resulting in a flat line. Meanwhile FFT (in blue) closely mimics the data, with both using $8$ values. 
    BOTTOM: The distribution of values for each dataset. SAX is built upon the assumption that the data follows Normal N(0,1) distribution (dotted in green). This is neither the case case for the PAA approximations nor the raw data.
	\label{fig:dataset_characteristics}
	}
\end{figure*}

Indexing is a widely used technique to accelerate similarity searches. Most indices for DS similarity search utilize summarized representations of the DS to map them into lower-dimensional spaces. Symbolic Aggregate approXimation (SAX)~\cite{shieh2008sax} is among the most popular techniques~\cite{isaxfamily}. 
SAX works by computing mean values over segments and then quantizing these values to form a SAX word. Its quantization employs a fixed alphabet of symbols, and the Normal Distribution is divided into equal-depth bins to derive these symbols. Consequently, SAX assumes that the DS are Normally distributed. The distance between two SAX words from DS A and B provides a lower bound to the ED between A and B, allowing for exact indexing using the GEMINI framework~\cite{FaloutsosRM94, Agrawal1993}, which will be explained in Section~\ref{sec:exact_sim_search_using_gemini}. The tighter the lower bound to the ED, the better the indexing performance, leading to reduced retrieval times. Although SAX's simplicity is sufficient for many scenarios, its rigid quantization scheme results in decreased performance when dealing with high-frequency data.

Figure~\ref{fig:dataset_characteristics} illustrates some cases where the SAX transformation is failing on our benchmark datasets (Section~\ref{sec:experiments}). The original signal is illustrated in black, and the mean values over $8$ segments in orange. For high-frequency datasets, using mean values (orange) fails to capture the signal's semantics, resulting in a flat line summarization (Figure~\ref{fig:dataset_characteristics} TOP). Additionally, the actual data distribution is often non-Gaussian, leading to poor lower bounding distances (LBDs) and degraded indexing performance (Figure~\ref{fig:dataset_characteristics} BOTTOM).

In this work, we present the SOFA (SymbOlic Fourier Approximation) index. It is based on two building blocks: SFA and a tree index inspired by MESSI and adapted to our approach. SOFA uses the Symbolic Fourier Approximation (SFA)~\cite{schafer2012sfa} for fast similarity search in large datasets, which addresses the aforementioned issues. SFA is based on a learned quantization of the Fourier transform. It allows for lower bounding the ED. Figure~\ref{fig:dataset_characteristics} illustrates the effectiveness of SFA in capturing the essence of a high-frequency DS. The SFA approximation using 4 Fourier coefficients (4 real+4 imag values) is depicted in blue. As evident from the figure, SFA provides a closer approximation to the actual DS compared to SAX, especially for signals with high variability. This improved representation capability and efficient computation, make SFA a powerful tool for similarity search. 
The MESSI index~\cite{peng2020messi,PengFP21} represents the state-of-the-art in-memory DS index. It makes use of the GEMINI-approach to answer exact similarity search queries. It was specifically developed for concurrent multi-threaded environments, modern hardware (SIMD), and for in-memory DS management. 

The contributions we make in this paper can be summarized as follows:
\begin{enumerate}
    \item We propose utilizing the learned symbolic summarization technique, Symbolic Fourier Approximation (SFA), for efficient DS similarity search. SFA demonstrates superior capability in capturing the latent semantics of high-frequency DS (Section~\ref{sec:experiments_ablation}).
    \item We introduce SOFA, an advanced indexing structure that synergistically combines SFA with a parallel tree index adapted from MESSI, a state-of-the-art DS index optimized for modern hardware architectures (Section~\ref{sec:framework}).
    \item In Section~\ref{sec:simd_lower_bounding_dist}, we present an efficient SIMD-based implementation of the SFA lower-bounding distance, which solves conditional branching and lower bounding.
    \item We present the most extensive benchmark to date for similarity search, encompassing 17 diverse datasets. It totals ~1TB of data, comprising 1,017,586,504 (1 billion) DS (Table~\ref{tab:datasets}).
    \item Our extensive experimental evaluation (Section~\ref{sec:experiments}) demonstrates that SOFA consistently outperforms existing methods. It is up to $38$ times faster than MESSI~\cite{peng2020messi}, and on average $2.5$ times faster. It is up to $10$ times faster than the a parallel sequential scan~\cite{RakthanmanonCMBWZZK12}, and $3-4$ times faster than FAISS~\cite{johnson2019billion}.    
\end{enumerate}

The remainder of this paper is structures as follows: In Section~\ref{sec:background} we review the background on similarity search. Section~\ref{sec:related_work} discussed the related work. Section~\ref{sec:framework} introduces the SOFA framework based on the MESSI index~\ref{sec:messi}, and the SFA summarization technique~\ref{sec:sax_summarization}. Section~\ref{sec:experiments} presents our experimental results and Section~\ref{sec:conclusion} concludes.

\section{Background and Definitions}\label{sec:background}

We define DS and the z-normalized Euclidean distance (z-ED), which we (like all prior works) will use throughout this work. 

\begin{definition}
\textbf{Data Series (DS)}: A data series $A=\left(p_1, \ldots , p_n \right)$ is an ordered sequence of $n$ points, where each point $p_i = (t_i, a_i)$ consists of a real value $a_i \in \mathbb{R}$ and a position $t_i \in \mathbb{N}$. 
\end{definition}

In subsequent discussion, we will disregard the positions and focus solely on the values. The order of values differentiate DS from vector data, such as embeddings learned from images or text. DS typically show low frequency, which allows for efficient summarization techniques to be applied. The later type of data is characterized by high variance in high frequency, which makes summarization more challenging. In the context of similarity search, the similarity of two subsequences is measured using the \emph{z-normalized ED}. 

\begin{definition}
\textbf{z-normalized Euclidean distance (z-ED)}: 
Given two univariate DS $A=(a_1, \ldots, a_l)$ with mean $\mu_A$ and standard-deviation $\sigma_A$ and $B=(b_1, \ldots, b_l)$ with $\mu_B$ and $\sigma_B$, both of length $l$, their (squared) z-ED is defined as:
$$\textit{d}(A, B)= \sqrt{\sum_{t=1}^{l}\left(\frac{a_t - \mu_A}{\sigma_A} - \frac{b_t - \mu_B}{\sigma_B}\right)^2}$$
\end{definition}

Given a query series $S_q$ of length $n$, a collection $\mathcal{S}$ of size $N$ consisting of DS of length $n$, similarity search aims to identify the nearest neighbor series $S_c \in \mathcal{S}$ for which the distance to $S_q$ is the smallest: $\forall S_o \in \mathcal{S}, S_o \ne S_c: d(S_c, S_q) \leq d(S_o, S_c)$

Similarity search can be accelerated using summarization techniques, which map DS into lower-dimensional space.

\begin{definition}
A \textbf{summarization} is a mapping $\mathcal{E} : \mathbb{R}^n \rightarrow \mathbb{R}^l$ that transforms a DS $S$ of length $n$ into a lower-dimensional series of length $l$, where $l \ll n$. 
\end{definition}

The lower bounding distance is a crucial property for similarity search, as it enables the exact retrieval of the nearest neighbors in a reduced $l$-dimensional space using the GEMINI-framework.

\begin{definition}
A \textbf{lower bounding distance (LBD)} $d'(\cdot, \cdot)$ for a distance measure $d(\cdot, \cdot)$ is a measure that satisfies the lower bounding property for DS $A$ and $B$:
$$d'(A', B') \leq \textit{d}(A, B)$$
\end{definition}

In this context, $A'=\mathcal{E}(A)$ and $B'=\mathcal{E}(B)$ represent the summarizations of $A$ $B$, respectively. (z-normalized) Euclidean LBD s have been defined for summarization techniques including iSAX~\cite{isaxfamily}, PAA~\cite{keogh2001dimensionality}, SFA~\cite{schafer2012sfa}, or DFT~\cite{Aggarwal1988}. 

\subsection{Exact Similarity Search using GEMINI}\label{sec:exact_sim_search_using_gemini}
Indexing high-dimensional DS is a considerable challenge due to the \textit{Curse of Dimensionality}~\cite{bellman1959adaptive}. As the dimensionality of the search space increases, the number of DS that must be examined typically grows exponentially. Consequently, executing an exact similarity query using spatial access methods (SAMs) can take longer than performing a sequential scan of all data~\cite{sprenger2018multidimensional}. 
SAMs traditionally become ineffective at dimensionalities ranging from $n=10$ to $20$ dimensions~\cite{FaloutsosRM94}. 

Research in~\cite{FaloutsosRM94, Agrawal1993} was the first to introduce the concept of dimensionality reduction of DS (summarization) prior to indexing. This approach, known as GEMINI, demonstrated that by using a lower-bounding distance measure on the summarization, queries in the reduced-dimensional search space return a super-set of the exact result as in the original search space. These false alarms are then pruned using the actual distance measure. 

The rationale behind GEMINI is to compute the lower-bounding distances for each DS $A$ relative to the query $Q$. The exact distance between $A$ and $Q$ is only calculated if the lower-bounding distance exceeds the current best-so-far exact distance computation. By providing a LBD, the Curse of Dimensionality is mitigated, reducing the effective dimensionality to $10-20$ dimensions of the summarization. 

Since the introduction of GEMINI, numerous summarization techniques with lower bounding techniques have been proposed, which can be categorized into real-valued and discrete-valued (symbolic) methods. Research on symbolic series is scarce. Symbolic methods combine dimensionality reduction with quantizing. SAX~\cite{lin2003symbolic} has established itself as the de-facto standard for exact similarity search on DS.

\subsection{Distance calculations using SIMD}

Single Instruction, Multiple Data (SIMD) is a parallel computing architecture designed to perform the same operation simultaneously on multiple data points, enabling efficient utilization of data-level parallelism~\cite{lomont2011introduction}. 
By leveraging SIMD, operation latency can be significantly as instructions are fetched only once and then executed in parallel across multiple data elements. Modern CPUs support SIMD vector widths of up to $512$ bits, allowing computations—such as floating-point or other $32$-bit data operations—to achieve speedups of up to $16$ times.
In the context of data series, SIMD has been effectively utilized for calculating Euclidean distance functions~\cite{tang2016exploit}. It has also been integrated into indexing techniques like ParIS+ and MESSI to optimize conditional branch calculations during lower-bound distance computations, enhancing performance in similarity search tasks~\cite{parisplus,peng2018paris,messi2,peng2020messi}.

\section{Related Work}\label{sec:related_work}
Since the introduction of the GEMINI framework~\cite{FaloutsosRM94} for exact similarity search on DS using lower bounding, a vast array of summarization techniques has emerged. While much of the foundational research in this field dates back several decades, its impact continues to resonate in modern data analysis.

The majority of research in this domain has focused on numeric techniques, which can be broadly categorized as follows: Piecewise Aggregate Approximation (PAA)~\cite{keogh2001dimensionality} uses mean values over fixed-length segments. Adaptive Piecewise Constant Approximation (APCA)~\cite{Chakrabarti2002} employs mean values over adaptive segments. Piecewise Linear Approximation (PLA)~\cite{Chen2007} uses line segments to represent DS. Chebyshev Polynomials~\cite{cai2004indexing} are another method for DS summarization. The Fourier Transform~\cite{Aggarwal1988,FaloutsosRM94} utilizes the frequency domain representation. Wavelets~\cite{popivanov2002similarity} are also used for DS summarization. Despite advancements in lower bounding strategies, there remains a notable gap in research focused on leveraging numerical techniques for indexing massive datasets. This is partly due to the substantial memory requirements of these methods, but also as these show no clear improvement over symbolic representations. In their study, \citet{schafer2012sfa} compared several techniques based on pruning power, namely, APCA, PAA, PLA, CHEBY, and DFT. They conclude that none outperformed DFT. Moreover, SFA consistently matched or exceeded the performance of all but DFT across nearly all scenarios. SFA is outperformed by DFT due to SFA's additional quantization step applied on top of DFT. The results that there is no significant difference were independently confirmed by~\citet{shieh2008sax}.
 
SAX~\cite{shieh2008sax} is a symblic summarization technique that is most commonly used for DS similarity search, as it has a much lower memory footprint requiring only a few bits (rather than a double/8 byte), and allows for lower bounding the ED, which is essential in finding the exact nearest neighbors~\cite{FaloutsosRM94}. 
The iSAX-family of indices is large~\cite{isaxfamily}, including iSAX~\cite{shieh2009isax}, iSAX2.0~\cite{Camerra2010}, iSAX2+~\cite{isax2plus}, ADS~\cite{zoumpatianos2016ads}, ADS+ and ADS-Full~\cite{zoumpatianos2016ads}, ParIS~\cite{peng2018paris}, ParIS+~\cite{parisplus}, MESSI~\cite{peng2020messi,PengFP21}, DPiSAX~\cite{DBLP:conf/icdm/YagoubiAMP17,dpisaxjournal}, ULISSE~\cite{ulisse,ulissejournal}, Coconut-Trie/Tree~\cite{DBLP:journals/pvldb/KondylakisDZP18}, Coconut-LSM~\cite{KondylakisDZP19}, Hercules~\cite{DBLP:journals/pvldb/EchihabiFZPB22}, and Dumpy~\cite{DBLP:journals/pacmmod/Wang0WP023}. 
MESSI~\cite{peng2020messi} is the SotA index for fast DS similarity search. Moreover, since all SAX-based indices~\cite{isaxfamily} use the same summarization technique, they will all benefit from the improvements introduced here.

The research on symbolic transformations in scars. SFA~\cite{schafer2012sfa} is a data-adaptive transformation based on the frequency domain. It is most common in dictionary-based classification approaches~\cite{middlehurst2024bake}. AN index was proposed based on SFA, named the SFA trie~\cite{schafer2012sfa}, inspired by building a prefix tree on the SFA word. However, it showed to be non-competitive to the SotA~\cite{echihabi2020lernaean}.

While other distance measures may be suitable in specific contexts, our focus is on Euclidean Distance (ED). Elastic distances, for instance, can be advantageous when working with limited data, such as in classification~\cite{shifaz2023elastic, middlehurst2024bake} or clustering tasks~\cite{holder2024review}. However, \citet{Shieh2008} demonstrated that the error rate of ED approaches that of Dynamic Time Warping (DTW) as the dataset size increases, rendering the difference negligible with a few thousand objects. Consequently, large-scale dataset indexing often favors ED due to its computational efficiency~\cite{evolutionofanindex}.

A similar, yet different area of research is on vector datasets~\cite{xie2023brief}. Other than DS, vector data has no ordering. I.e. the values may be re-ordered in any arbitrary order. This can result in very high variance in high frequencies, which makes the application of summarization techniques hard. Thus, most research in vector data focuses on approximate retrieval~\cite{tian2023approximate}. FAISS~\cite{johnson2019billion} is a SotA approach for exact similarity search on vector datasets. In this work, we compare to the FAISS approach.
Approximate similarity search~\cite{johnson2019billion} may not always yield exact results, but results that are very close. 

MASS~\cite{zhong2024mass} and the UCR suite~\cite{dau2019ucr} were originally designed for subsequence search, targeting the problem of finding the lowest distance subsequence in a long data series. Our paper, however, focuses on whole-series matching. MASS is less effective and up to $5$ times slower than the UCR suite for this task (see Fig 3 in~\cite{lernaeanhydra2}). Thus, we did not include MASS in our experiments, but compare to the UCR suite.

\section{SOFA - Fast, Exact Sim. Search}\label{sec:framework}

\textit{SOFA (SymbOlic Fourier Approximation)} is based on two building blocks: a tree index adapted from MESSI and the SFA symbolic summarization. 
In the following section, we first introduce MESSI for answering exact queries using the GEMINI approach for exact similarity search. From Section~\ref{sec:sax_summarization}, we present the SAX summarization, and compare it to SFA, and finally sketch how to implement it within MESSI (Section~\ref{sec:messi_sfa}).

\subsection{MESSI}\label{sec:messi}


MESSI~\cite{peng2020messi,PengFP21} represents the state-of-the-art in-memory DS indexing~\cite{isaxfamily}. It was specifically designed to exploit concurrent multi-threaded environments and for in-memory DS management for exact similarity search using the GEMINI framework. 
It utilizes variable cardinalities for iSAX summaries, varying the number of bits used for each segment's symbols, to build a hierarchical index. 

\subsection{Structure}
This tree consists of three types of nodes:
\begin{enumerate}
\item \textbf{Root Node:} Points to multiple child nodes (up to 2w in the worst case, where w is the number of bits), representing all possible iSAX summaries.
\item \textbf{Inner Nodes:} Each has two children and holds an iSAX summary representing all series in its subtree.
\item \textbf{Leaf Nodes:} Store iSAX summaries of several DS and pointers to these series.
\end{enumerate}
When the number of series in a leaf node exceeds its capacity, the leaf splits into two new leaves, becoming an inner node. This split is achieved by increasing the cardinality of one segment's iSAX summary to ensure a balanced distribution of series between the new leaves~\cite{isax2plus,zoumpatianos2016ads}). The iSAX summaries for the new leaves are then adjusted by setting the new bit to $0$ and $1$, respectively.

\subsection{Exact Similarity Search} 
For query answering, MESSI first performs an Approximate Search by traversing the index tree to find the best candidate series.  Then, it computes an approximate answer by calculating the real distance,  called Best-So-Far (\textit{BSF}), between the query series and the candidate series.  The leaf with the smallest lower-bound distance to the query is identified,  and \textit{BSF} is used to prune as many candidate series as possible from the dataset during the tree parsing.

MESSI employs multiple \textit{index workers} to traverse index subtrees in parallel,  using \textit{BSF} to determine which subtrees can be pruned. Each subtree is assigned to a single worker to minimize synchronization needs.  Each worker only needs to synchronize during subtree selection.  The leaves of unpruned subtrees are placed into a fixed number of priority queues, ordered by the lower-bound distance between the PAA of the query and the iSAX summary of the leaf node. 

The workers access these priority queues using locks. Each index worker processes a priority queue by repeatedly calling \textit{DeleteMin()} to retrieve a leaf node. The worker checks if the lower-bound distance between the query and the leaf exceeds the current \textit{BSF}. If it does, the leaf is pruned. Otherwise, the worker examines the series within the leaf by computing the lower-bound distance between the PAA of the query and the iSAX summaries of each series. If this lower-bound distance is less than the \textit{BSF}, the worker calculates the real distance using the raw values further. If a series with a smaller distance than the current \textit{BSF} is found, the \textit{BSF} is updated.

When a worker encounters a node with a distance greater than the \textit{BSF}, 
it abandons the current priority queue and selects another, as all remaining elements in the abandoned queue have higher distances and can be pruned. 
This process continues until all priority queues are processed, and the \textit{BSF} is updated accordingly. The final value of the \textit{BSF} is returned as the query result. Note that the above process is trivially extended to return the $k$-NN. For further details, readers are referred to~\cite{peng2020messi,PengFP21}. 


\subsection{iSAX - A Static Symbolic Representation}\label{sec:sax_summarization}
The indexable SAX (iSAX) summarization technique combines (a) the Piecewise Aggregate Approximation (PAA) with (b) a fixed equal-depth binning quantization derived from the Normal distribution N(0,1). iSAX transforms a DS into a word as follows:
\begin{enumerate}
    \item \textbf{Segmentation:} The DS is divided into $l$ segments of equal length $n/l$.
    \item \textbf{Aggregation:} Each segment is represented by its mean value, also referred to as Piecewise Aggregate Approximation (PAA)~\cite{keogh2001dimensionality}. This step acts as a low pass filter, as high variance segments, like high-frequency segments, dropouts or noise, of a series are smoothed. 
    \item \textbf{Fixed Quantization:} The $l$ segment (PAA) means are mapped 
    to symbols using breakpoints that assume a Normal distribution of the data.   
\end{enumerate}

Quantization in iSAX is build upon the assumption that values sampled from DS follow a Normal distribution $N(0,1)$, with mean $0$ and std $1$. iSAX ignores the actual data distribution and applies equal-depth binning applied to the Normal distribution, given the fixed-size alphabet $\Sigma$. These bins are typically hard coded. 
SAX has two parameters: (a) the length $l$ and (b) the number of symbols $|\Sigma|$. In practice, the number of symbols is a small number, with as few as $256$ symbols, which can be represented by $8$ bits.  The number $256$ is based on an $8$-bit char accommodating $256$ states, optimizing space use. More symbols would require $16$-bits, doubling space, or slow bit operations, with negligible TLB efficiency gains (see Figure 6 of~\cite{Camerra2010}).

The transformation of a single series into a word takes $\mathcal{O}(n + l \times |\Sigma|)$
SAX makes explicit assumptions about the data and may lose information about high-frequency components or fine-grained details due to the averaging process. SAX provides a LBD to the ED called \textit{mindist}~\cite{shieh2008sax}.

\subsection{SFA - A Learned Symbolic Representation}

The Symbolic Fourier Approximation (SFA) combines (a) the Discrete Fourier Transform (DFT) with (b) feature selection and (c) binning of the actual distribution of the Fourier coefficients. SFA transforms a DS into a word as follows:
\begin{enumerate}
    \item \textbf{Transformation:} A DS is transformed into the frequency domain using the DFT.
    \item \textbf{Feature Selection:} Only a subset of $l < n$ Fourier values (real or imaginary part) is retained, as these capture the main structure of the data.
    \item \textbf{Learned Quantization:} Each Fourier value is discretized into a symbol using equi-depth or equi-width binning based on the actual distribution of the frequency spectrum, i.e., the real or imaginary values.
\end{enumerate}

SFA has two parameters: (a) the length $l$ and (b) the number of symbols $|\Sigma|$. For SFA the same default values apply as for SAX.

SFA is not based on any assumptions, and learns bins from the actual data distribution in Frequency domain. Yet, it can be more computationally intensive due to the Fourier transform with $\mathcal{O}(n \log n + l \times |\Sigma|)$ for each series, and binning based on actual data distribution.

\begin{figure}[t]
    \centering  
	\includegraphics[width=1.00\columnwidth]{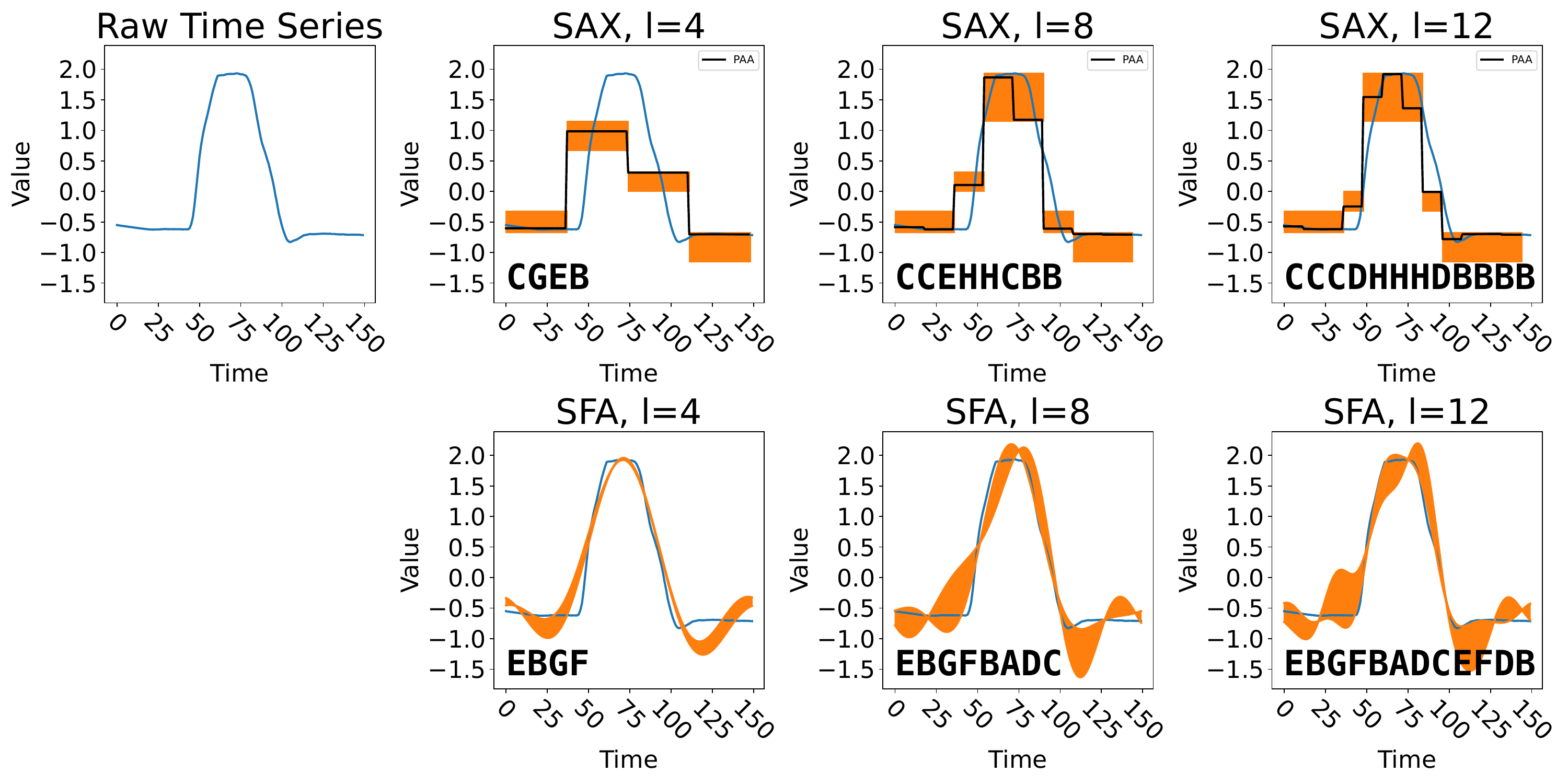}
	\caption{
	The figure illustrates the summarization of a DS using SAX (top) and SFA (bottom), both employing an 8-symbol alphabet ('a' to 'h') with $4$ to $12$ values. SAX generates a staircase-like envelope around the raw signal (shown in orange). In contrast, SFA constructs an envelope around the Fourier transform, closely approximating the original signal.
	\label{fig:sax_vs_sfa}
	}
\end{figure}

Figure~\ref{fig:sax_vs_sfa} illustrates the differences between both approaches, when using the same number of symbols and lengths $l\in[4,8,12]$ each. SAX gives a staircase like approximation of a DS, which for lower word lengths fails to capture the intrinsic shape of the signal. SFA represents a smooth envelop around the Fourier coefficients of the DS.

\subsubsection{Transformation and Learned Quantization}\label{sec:transform_and_learning}

The quantization in SFA is learned from the distribution of the real or imaginary Fourier values of the transformed series using a technique called \textbf{Multiple Coefficient Binning (MCB)}. The goal of MCB in SFA is to minimize the loss of information introduced by quantization. A better representation of the original signals enhances pruning efficiency during query execution. 

SFA learns $l$ sets of quantization intervals using MCB, one for each real or imaginary Fourier value. First, all $N$ series are transformed using the Discrete Fourier Transform (DFT). Then $l$ real and imaginary Fourier values are selected using a feature selection strategy (Section~\ref{sec:feature_selection}. From one (real or imaginary) Fourier value a set of $\alpha=|\Sigma|$ bins is learned using an alphabet  $\Sigma$. As a result of the $j$-quantization, the following bins are derived: 
$$\left[\beta_{j}(a-1),\beta_{j}(a)\right), \textit{for }j\,\epsilon\,\left[0\ldots l\right),\,a\,\epsilon\,\left[1\ldots |\Sigma|\right)$$

We label bins by assigning the $a$-th symbol of alphabet $\Sigma$ to it: 
$$\textit{symbol}^{(a)} \equiv \left[\beta_{j}(a-1),\beta_{j}(a)\right), \textit{for }j\,\epsilon\,\left[0\ldots l\right),\,a\,\epsilon\,\left[1\ldots |\Sigma| \right)$$

When proposed, equi-depth binning was used~\cite{schafer2012sfa}. However, to achieve a tighter lower bound to the ED, it is essential to maximize the size of each interval. Equi-width proves to be superior, as it generates uniformly sized bins which are equally large, thus enhancing the accuracy of the lower bound (see Section~\ref{sec:experiments_exact_search}).

\begin{figure*}[t]
    \centering  
	\includegraphics[width=1.80\columnwidth]{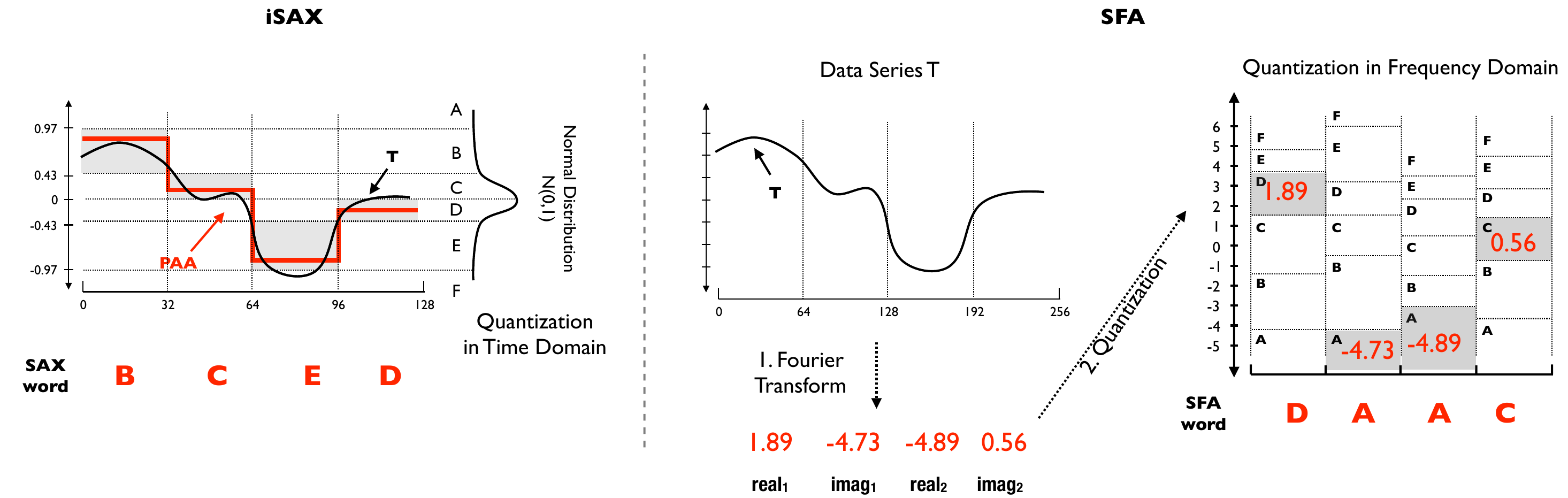}
	\caption{
	The figure illustrates the two summarization techniques iSAX (left) and SFA (right). SAX aggregates the DS over intervals using PAA, and quantizes the mean values into symbols BCED using bins derived from equi-depth binning the Gaussian distribution. SFA transforms the DS into frequency domain, and separately quantizes the real and imaginary values into symbols using learned bins.
	\label{fig:quantization_sax_vs_sfa}
	}
\end{figure*}

Figure~\ref{fig:quantization_sax_vs_sfa} (right) illustrates the process of transforming a series into a word. iSAX (left) uses the same set of intervals based on the Normal distribution for all segment means, resulting in words, like \textit{BCED}. SFA (right) first transforms the series into real and imaginary Fourier values, selects those with the highest variance, and learns different sets of bins for each value. These bins are then used to transform the series into words, like \textit{DAAC}.

\subsubsection{Novel Feature Selection}\label{sec:feature_selection} Commonly, for SFA the first Fourier values are retained, which acts as a low-pass filter. 
However, this approach reduces high variance components of the DS, which leads to degenerated performance for lower bounding. To address this limitation, we introduce a novel \emph{variance-based} Fourier value selection strategy. The rationale behind this strategy is that maximizing the LBD requires maximizing the width of the quantization bins (intervals).

Given a dataset of DS D, the variance of each Fourier coefficient, we select those $l$ coefficients of each transformation with highest variance: $$\textsc{best}=\textsc{k-argmax} \left( VAR(DFT\left( D \right), axis=1) \right)$$

This strategy begins by computing the variance of each real and imaginary component of the Fourier transform for the $N$ DS. It then selects the top $l$ real and imaginary components with the highest variance. 

The reason a larger variance improves the LBD is that it allows for wider quantization bins. Wider intervals capture more variability in the data, leading to a more accurate and informative representation, which in turn enhances the pruning efficiency during query execution. We will show the impact of this selection strategy in our ablation studies  (Section~\ref{sec:experiments_ablation}).

\subsubsection{ED Lower Bounding Distance}\label{sec:lower_bounding_distance}

Given the DFT representations $DFT(A)=A'=(a'_{0},\ldots,a'_{l-1})$
of DS $A$ and $DFT(B)=B'=(b'_{0},\ldots,b'_{l-1})$ of DS $B$ the \emph{DFT LBD} to the ED $d_{ED}$ is defined as~\cite{rafiei98efficient} (for $l < n/2$):
\begin{equation}
d_{DFT}^{2}(A',B')=(a'_{0}-b'_{0})^{2}+2\sum_{i=1}^{l-1}(a'_{i}-b'_{i})^{2}\leq d_{ED}^{2}(A,B)\label{eq:DFT}
\end{equation}

Note that the first DFT coefficients $a'_{0}$ and $b'_{0}$ represents the mean value, which is $0$ for z-normalized DS, thus can be omitted. The \emph{SFA Euclidean LBD} between a DFT representation
$DFT(B)=B'=(b'_{0},\ldots,b'_{l-1})$ and an SFA representation $SFA(a)=A'=(a'_{0},\ldots,a'_{l-1})$ is calculated by exchanging the pairwise difference of the numerical values in Equation~\ref{eq:DFT} by a \emph{$dist_{i}$} function,
which measures the distance between the $i$-th symbol and the $i$-th numerical value~\cite{schafer2012sfa}:
\begin{align*} \label{eq:dftsax2}
d_{SFA}^{2}(A',B') &= mind_{0}(a'_{0},b'_{0})^{2}+2\sum_{i=1}^{l-1}mind_{i}(a'_{i},b'_{i})^{2} \\
d_{SFA}^{2}(A',B') &\leq d_{ED}^{2}(A,B)
\end{align*}

Again, the first term is 0 in the case of z-normalized DS and can be omitted. The distance $dist_{i}$ between a numerical value $b'_{i}$ and symbol $a'_{i}$, represented by its lower and upper breakpoints $a'_{i}=\left[\beta_{i}(a-1),\beta_{i}(a)\right)$, is defined as the distance to the lower breakpoint if $b'_{i}$ is smaller or the upper quantization breakpoint if $b'_{i}$ is larger:
\begin{equation}
mind_{i}(a'_{i},b'_{i})\equiv\begin{cases}
0, & \textrm{if}\,b'_{i}\,\epsilon\,\left[\beta_{i}(a-1),\beta_{i}(a)\right)\\
\beta_{i}(a-1)-b'_{i}, & \textrm{if}\,b'_{i}<\beta_{i}(a-1)\\
b'_{i}-\beta_{i}(a), & \textrm{if}\,b'_{i}>\beta_{i}(a)
\end{cases}\label{eq:dist_i}
\end{equation}

\begin{figure}[th]
    \centering  
	\includegraphics[width=1.00\columnwidth]{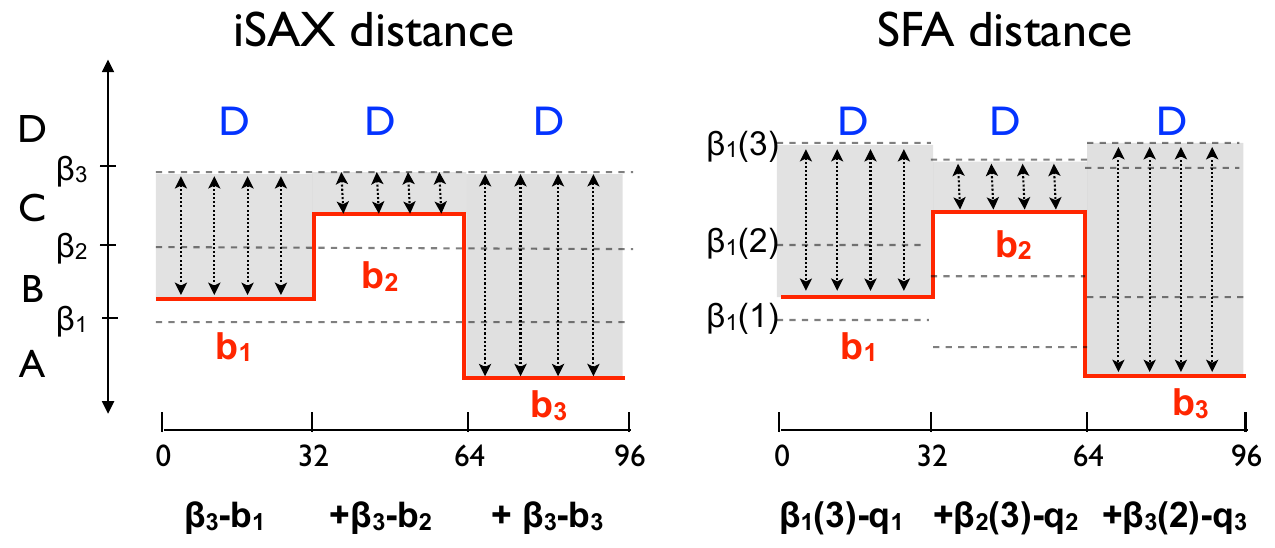}
	\caption{
	A comparison of the iSAX (left) and SFA (right) Euclidean LBD. iSAX uses the same fixed break points for each PAA value. SFA uses learned break points for each Fourier value (mean or imaginary values).
	\label{fig:lower_bounding_distances}
	}
\end{figure}

Figure~\ref{fig:lower_bounding_distances} (right) illustrates the SFA $mind_{i}$ in contrast to the iSAX lower bounding definition (left). For both the distance between the word $A'=DDD$ and $B'=(b_1, b_2, b_3)$ are computed. SFA uses learned bins per Fourier value, as opposed to the one set of fixed intervals used in iSAX.

\subsection{Pseudo-Code}\label{sec:pseudocode}

The pseudo-code for SFA is given in Algorithms~\ref{alg:mcb} and ~\ref{alg:sfa_transform}. MCB is used to learn quantization bins and the best Fourier coefficient indices. SFA transform is used to transform a single data series using the learned bins.

\begin{algorithm}[t]
\caption{MCB Quantization}\label{alg:mcb}
\begin{algorithmic}[1]
\Require A set X, of data series of length $n$, with $|X| = N$, the number of coefficients $l$ (default $16$), the alphabet size $a$ (default $256$), the sampling ratio $r$ (default $1\%$)
    
    \vspace{0.2cm}

\State \textit{Step 1: Sampling and Discrete Fourier Transform}
    \State $X_{subsample} = \textit{sample}(X, r)$
    \State $X_{DFT} = \textit{DFT}(X_{subsample})$
    \vspace{0.2cm}

\State \textit{Step 2: Determine $l$ best coefficient indices by variance}
    \State $\textsc{best\_l}=\textsc{k-argmax} \left( VAR(X_{DFT}, axis=1), k=l \right)$
    \State $X_{best} = X_{DFT}[:, best\_l]$
    \vspace{0.2cm}

\State \textit{Step 3: Learn $l$-sets of bins using equi-width binning}
    \State $\textsc{bins} = \textit{array of size }  (l \times a)$ 
    \For{$j = 1$ to $l$}
        \State $\textsc{bins[j]} = \textsc{apply-equi-width}(X_{best}[j], a)$
    \EndFor
    \vspace{0.2cm}

\State \Return $\textsc{bins}$, $\textsc{best\_l}$
\end{algorithmic}
\end{algorithm}

MCB (Algorithm~\ref{alg:mcb}) begins by sub-sampling the dataset using a sampling ratio $r$ (line~2), with a default value of $1\%$. The impact of varying $r$ is explored in the experiments (Section~\ref{sec:subsampling}). The subsample is then transformed using the Discrete Fourier Transform (line~3). Subsequently, the Fourier coefficients are ranked by selecting the indices that maximize variance across the subsample (line~5). These top coefficients are retained for further processing (line~6). Finally, $l$ sets of bins, each containing $a$ bins, are learned using either equi-width or equi-depth binning (lines~8–11). The algorithm returns the optimal bin boundaries and selected Fourier coefficient indices.

\begin{algorithm}[t]
\caption{SFA Transform}\label{alg:sfa_transform}
\begin{algorithmic}[1]
\Require A single data series $T$ of length $n$, quantization bins $\textit{bins}$ of size $(l \times a)$, Fourier coefficient indices $\textit{best\_l}$
\vspace{0.2cm}

\State \textit{Step 1: Transform T keeping only best $l$ coefficients}
    \State $T_{DFT} = \textit{DFT}(T)$
    \State $T_{best} = T_{DFT}[best\_l]$
    \vspace{0.2cm}

\State \textit{Step 2: Apply quantization}    
    \State $T_{SFA} = \textsc{apply-quantization}(T_{best}, bins)$
    \vspace{0.2cm}

\State \Return $T_{SFA}$
\end{algorithmic}
\end{algorithm}

The SFA transform (Algorithm~\ref{alg:sfa_transform}) operates on a single data series using the pre-learned breakpoints (using MCB) and selected Fourier coefficient indices. First, the DFT is applied to the data series (line 2), and only the top-ranked coefficients are retained (line 3). Finally, these coefficients are quantized based on the learned breakpoints to generate words (line 5).

\subsection{SOFA index}\label{sec:messi_sfa}
\begin{figure}[t]
    \centering  
    \hspace*{-0.3cm}
	\includegraphics[page=1,width=1.1\columnwidth]{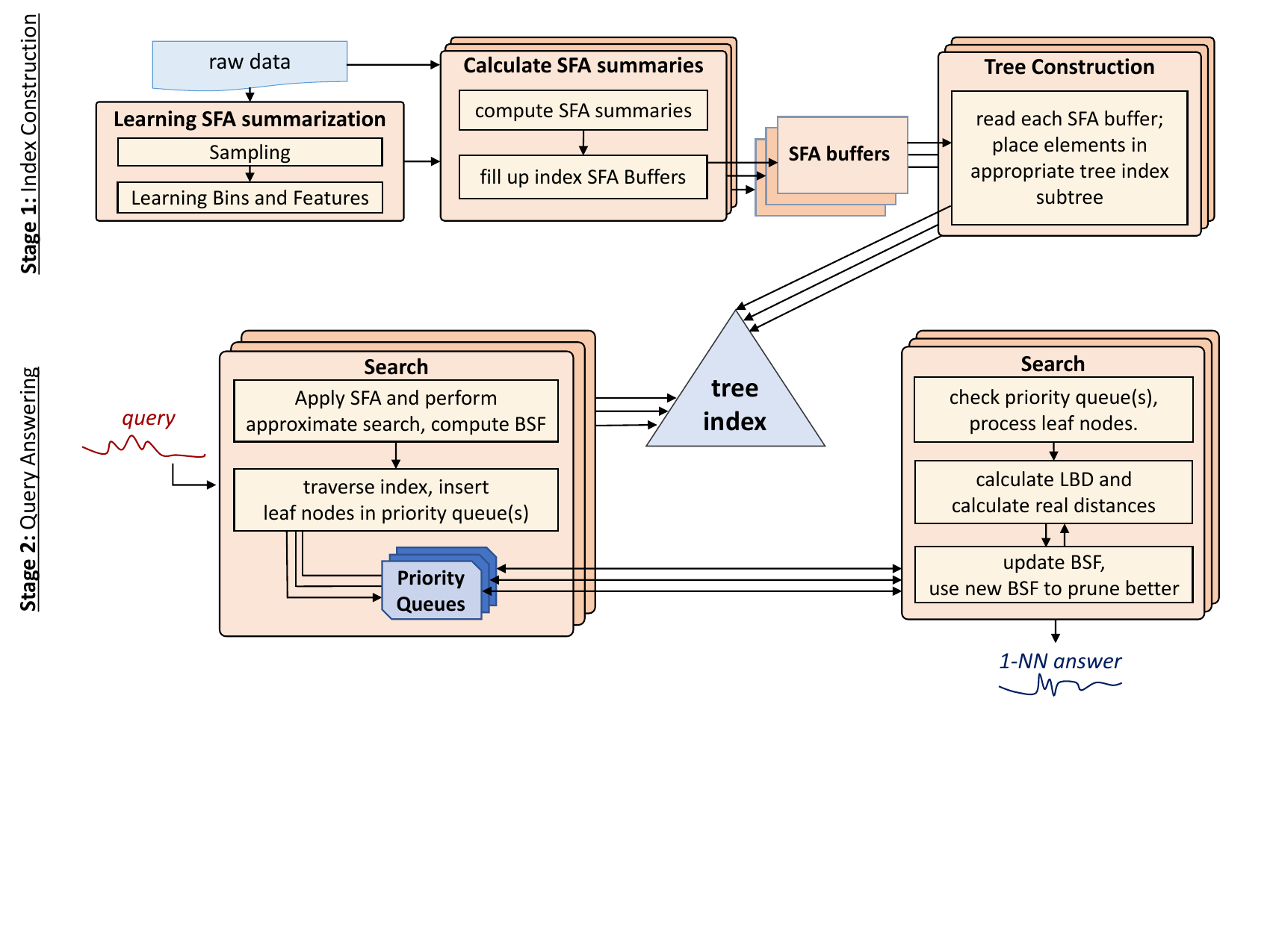}
    \vspace*{-2.4cm}
	\caption{
	Workflow of SOFA for exact similarity search. First, a fraction of the DS is sampled and Fourier transformed. Bins are learned, and the best Fourier coefficients selected. Using the learned transformation, all DS are transformed to create the index. To answer a query, the query DS is SFA transformed, and the MESSI-based index is used to retrieve the exact 1-NN using the SFA lower bound.
	\label{fig:workflow}
	}
\end{figure}

The \textit{SOFA (SymbOlic Fourier Approximation)} index is based on the MESSI index and the SFA representation.  It allows for fast and exact similarity search is illustrated in Figure~\ref{fig:workflow}. 

The process begins by sampling a fraction ($1\%$) of the DS to learn the SFA summarization with alphabet size $256$ (compare Section~\ref{sec:sax_summarization}). The sample undergoes a Fourier transformation, followed by the learning of quantization bins from the Fourier values (Section~\ref{sec:transform_and_learning}). 
Then, $8$ learned Fourier coefficients are selected based on highest variance, which is equal to $16$ imag and real float values. Utilizing the learned SFA, all DS are transformed and indexed. To answer a query, the query is first transformed using SFA, and then the index is used to search exact nearest neighbors using  GEMINI (Section~\ref{sec:exact_sim_search_using_gemini}), and the SFA lower bound (Section~\ref{sec:lower_bounding_distance}). 



\subsection{SOFA LBD calculation using SIMD}\label{sec:simd_lower_bounding_dist}

\begin{figure}[t]
    \centering  
    \hspace*{-0.5cm}
	\includegraphics[width=1.1\columnwidth]{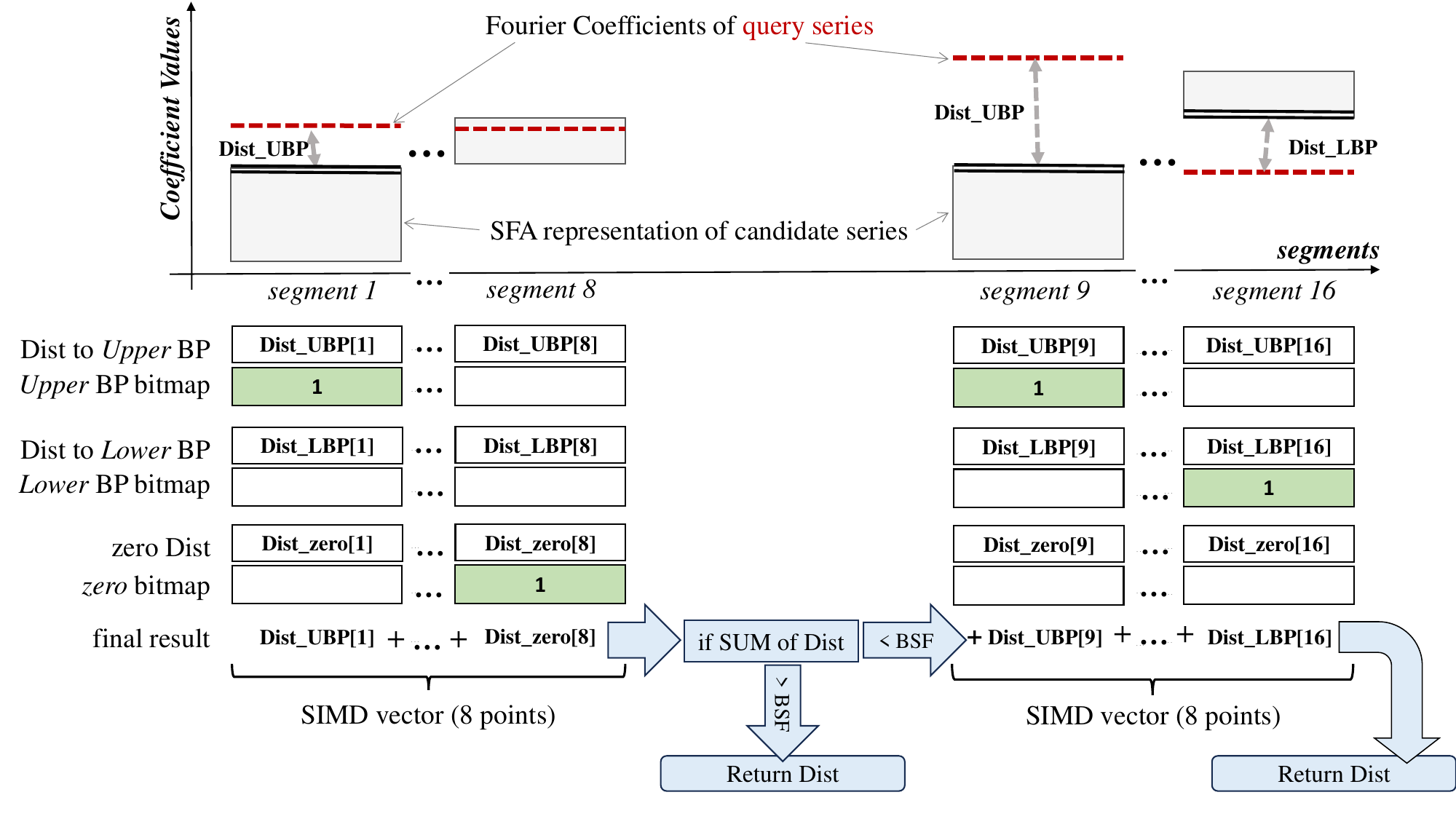}
	\caption{
	SIMD lower bound distance calculation illustration using bitmaps for conditional branching, and chunks for early abandoning.
	\label{fig:SIMD}
	}
\end{figure}

Similar to MESSI, SOFA uses SIMD to accelerate both lower-bound and real distance calculations. Algorithm~\ref{alg:min_dist_fft_sfa} and Figure~\ref{fig:SIMD} illustate the process of lower bound distance (LBD) computations between a query's DFT coefficients and an SFA word of an indexed series using SIMD. Using SIMD enables parallel computation of $8$ or $16$ distances in a single instruction, utilizing $256$-bit or $512$-bit vectors with eight $32$-bit floating-point elements each.

The SIMD implementation addresses two key challenges: (i) optimizing branching based on upper, lower, and zero bounds, and (ii) enabling early termination while utilizing SIMD registers.

\paragraph{Conditional Branching} Unlike simpler branching techniques~\cite{tang2016exploit}, LBD computations require distinct conditional branches and specific assignments for each SIMD vector position. 
To compute the LBD between the DFT coefficients of the query series and an SFA word, three conditions are evaluated: (i) the DFT coefficient lies above the \textit{UPPER} breakpoint of the corresponding SFA representation of the candidate series (quantization interval), (ii) below the \textit{LOWER} breakpoint, or (iii) within the SFA quantization interval, denoted as \textit{ZERO} (compare Figure~\ref{fig:lower_bounding_distances} and Eq.~\ref{eq:dist_i}). 
All three conditions have to be processed simultaneously across SIMD positions, and a conditional mask is applied to extract the correct result.

\paragraph{Early Abandoning}
The Fourier coefficients in SFA are prioritized by relevance (compare Section~\ref{sec:feature_selection}), with high-variance coefficients contributing more to the overall distance between two data series. While SIMD processes calculations in parallel batches, it cannot inherently leverage early abandoning after each computation.

Figure~\ref{fig:SIMD} provides an overview of the complete SIMD LBD computation process. Initially, the distance between each query's Fourier coefficient and the SFA word's intervals is calculated (top) using Eq.~\ref{eq:dist_i}. The query is then processed in chunks of 8 points (bottom). After processing each chunk, the current distance is compared to the BSF, and if it exceeds the BSF, the current distance is returned. To ensure efficient branching, \textit{UPPER}, \textit{LOWER}, and \textit{ZERO} conditions are evaluated using bitmaps, and aggregated into the final distance. 
Fully vectorizing all calculations enhances the speed of distance computations while minimizing the overhead of switching between vector and scalar registers, which operate on separate hardware. 

\paragraph{Pseudocode}

\begin{algorithm}[t]
\caption{SIMD-Optimized Min Distance Calculation}\label{alg:min_dist_fft_sfa}
\begin{algorithmic}[1]
\Require quantization bins $\textit{bins}$, FFT coefficient representation of query data $F_Q$, SFA representation of candidate $S_C$, BSF $bsf$.
\State Initialize vectors $V_{F\_Q}$,$V_{S\_C}$, $V_D$, $i=0$, $D_C = 0$
    \For{$i$th SIMD block size data}
        \State $V_{ F\_Q} \leftarrow F_Q$[$i$]
        \State $V_{ S\_C} \leftarrow S_C$[$i$]
        \State $V_{B\_U}$, $V_{B\_L} \leftarrow$  Gather\_bound $(V_{ S\_C}$, $bins)$
        \State $V_{D\_U}$, $V_{D\_Z}$, $V_{D\_L}$ $\leftarrow$ Caldist$(V_{ F\_Q}$, $V_{B\_U}$, $V_{B\_L})$
        
        \State $V_{M\_U}$, $V_{M\_Z}$, $V_{M\_L} \leftarrow$ Genmask$(V_{ F\_Q},V_{B\_U},V_{B\_L})$
        \State $V_D \leftarrow (V_{D\_L} and V_{M\_L}) or (V_{D\_U} and V_{M\_U})$

        \State $D_C \leftarrow Sum( V_D)$
        \If{$D_C$ exceeds $bsf$}
            \State \Return $D_C$
        \EndIf
        \State $i++$
    \EndFor
\State \Return $D_C$
\end{algorithmic}
\end{algorithm}

To address \emph{early abandoning}, Algorithm~\ref{alg:min_dist_fft_sfa} splits the coefficients into smaller chunks of up to $8$ data points for $256$ vector size (lines~2-14), enabling SIMD acceleration. 
After processing each chunk, intermediate results are aggregated, allowing early abandoning if the cumulative sum exceeds the best-so-far distance (lines~10-12). 
This approach combines the speed of SIMD with the efficiency of early abandoning, optimizing the balance between performance and computational savings.

To address \emph{conditional branching}, Algorithm~\ref{alg:min_dist_fft_sfa} generates three branch masks (\textit{UPPER}, \textit{LOWER}, and \textit{ZERO}), with each mask containing a value of \textit{1} at positions in the SIMD vector where the corresponding condition is met (line~7). 
For instance, if the first segment of the query lies above the interval of the word representation, the \textit{UPPER} mask is set to \textit{1} for that position. 
The distance value for the \textit{UPPER} branch is then selected for that position. 
Using SIMD instructions (e.g., AVX, AVX2, SSE3)\cite{coorporation2009intel}, these masks are generated efficiently (lines~7). 
A logical \textit{AND} operation is applied between each branch result and its corresponding mask, setting irrelevant branch results to zero (lines~8). 
Finally, the results from all branches are combined into a single vector, retaining only the correct values for each position and completing the distance computation.


\section{Experimental Evaluation}\label{sec:experiments}

In this section, we present a threefold experimental evaluation on $17$ real datasets, with a total of 1 billion DS or ~1TB of data. First, we will outline our setup, competitors and datasets. Then, we show the results of the exact similarity search comparing MESSI, SOFA to FAISS (Section~\ref{sec:experiments_exact_search}). 
We perform an ablation study in Section~\ref{sec:experiments_ablation}.


\paragraph{Competitors}
We compared \textbf{SOFA} with \textbf{MESSI}, \textbf{UCR Suite-P}~\cite{RakthanmanonCMBWZZK12}, and \textbf{FAISS IndexFlatL2} ~\cite{johnson2019billion}. UCR Suite-P is a parallel implementation of the state-of-the-art optimized serial scan technique using SIMD. In UCR Suite-P, each thread is allocated a segment of the in-memory DS array, allowing all threads to concurrently and independently process their assigned segments. The real distance calculations are performed using SIMD, and synchronization occurs only at the end to compile the final result. For FAISS, we used the CPU implementation of IndexFlatL2, the exact  similarity search index under L2 (ED).

The design principle for MESSI and SOFA involves sequential query processing, handling queries one after another. This approach simulates an exploratory analysis scenario where users generate new queries based on the results of previous ones. FAISS, however, cannot leverage parallelism within single query processing. 

Therefore, to take advantage of FAISS's capabilities, we process queries in mini-batches equal to the number of available cores, which exploits the inherent embarrassingly parallel nature of batch processing. FAISS (IndexFlatL2) was installed with Intel MKL support, and the number of OMP threads was configured according to the available cores in each experiment. 
The code, scripts, and notebooks for all algorithms used in this paper are available online~\cite{messi2}.

\paragraph{Setup}
We used a server with 2x Intel Xeon 6254 3.1Ghz CPUs (18 cores each, and 36 cores in total) and 756GB RAM. All algorithms were implemented in C, and compiled using GCC v7.5.0 on OpenSuse Linux v15.5. As we have $36$ cores available, we performed experiments with $9$, $18$, $36$ cores, each. For both MESSI and SOFA, we used consistent configurations across all experiments: `initial-lbl-size`, `min-leaf-size`, and `leaf-size` were set to $20000$, with the queue size matching the number of cores. The alphabet size was set to $256$ and the word length to $16$ for both SAX and SFA. SFA quantization was learned using a fraction of $1\%$ of the dataset, and selected the highest variance Fourier values (real or imaginary) from the first $16$ Fourier coefficients. As we test the batch case, where all data is available, we chose $1\%$ to avoid overfitting the train dataset, which could lead to decreased performance on the independent query data.


\paragraph{Datasets}
Table~\ref{tab:datasets} provides a summary of the properties of the $17$ datasets utilized in this study. They originate from various fields, including seismology (ETHZ, Iquique, LenDB, NEIC, OBS, SCEDC), astronomy (Astro), neuroscience (SALD), and vector datasets (Deep1B, BigANN, SIFT1b). Among them, only five datasets (Astro, BigANN, Deep1B, SALD, SIFT1b) have been previously used in studies on similarity search. Each is accompanied by a distinct set of $100$ queries. The remaining $12$ datasets are seismic datasets sourced from~\citet{woollam2022seisbench}. The samples were obtained by segmenting the seismic data into non-overlapping windows. Queries were generated using only the location of the primary (P)-wave of a seismic event, which travels faster than the secondary (S)-wave. The dataset sizes vary from 500k to 100M series, with series lengths ranging from 96 to 256 floats. Overall, there are 1 Billion real DS, corresponding to ~1TB on disk. In line with previous studies, we used distinct sets of $100$ query series for each dataset, ensuring they were kept separate from the indexed data.

\begin{table}
\centering
\caption{Characteristics of the $17$ datasets used}
\begin{NiceTabular}{c|c|c}
\toprule 
Dataset Name & \# of Series & Series Length\tabularnewline
\midrule 
\textbf{Astro}~\cite{soldi2014long} & 100,000,000 & 256\tabularnewline
\textbf{BigANN}~\cite{simhadri2022results} & 100,000,000 & 100\tabularnewline
\textbf{Deep1b}~\cite{babenko2016efficient} & 100,000,000 & 96\tabularnewline
\textbf{ETHZ}~\cite{woollam2022seisbench} & 4,999,932 & 256\tabularnewline
\textbf{Iquique}~\cite{woollam2019convolutional} & 578,853 & 256\tabularnewline
\textbf{ISC\_EHB\_DepthPhases}~\cite{munchmeyer2024learning} & 100,000,000 & 256\tabularnewline
\textbf{LenDB}~\cite{magrini2020local} & 37,345,260 & 256\tabularnewline
\textbf{Meier2019JGR}~\cite{woollam2022seisbench} & 6,361,998 & 256\tabularnewline
\textbf{NEIC}~\cite{yeck2021leveraging} & 93,473,541 & 256\tabularnewline
\textbf{OBS}~\cite{bornstein2024pickblue} & 15,508,794 & 256\tabularnewline
\textbf{OBST2024}~\cite{niksejel2024obstransformer} & 4,160,286 & 256\tabularnewline
\textbf{PNW}~\cite{ni2023curated} & 31,982,766 & 256\tabularnewline
\textbf{SALD}~\cite{url:SALD} & 100,000,000 & 128\tabularnewline
\textbf{SCEDC}~\cite{center2013southern} & 100,000,000 & 256\tabularnewline
\textbf{SIFT1b}~\cite{jegou2011searching} & 100,000,000 & 128\tabularnewline
\textbf{STEAD}~\cite{mousavi2019stanford} & 87,323,433 & 256\tabularnewline
\textbf{TXED}~\cite{chen2024txed} & 35,851,641 & 256\tabularnewline
\bottomrule
\end{NiceTabular}\label{tab:datasets}
\end{table}


\subsection{Index Creation}\label{sec:index_creation}
\begin{figure}[t]
    \centering  
	\includegraphics[width=1.00\columnwidth]{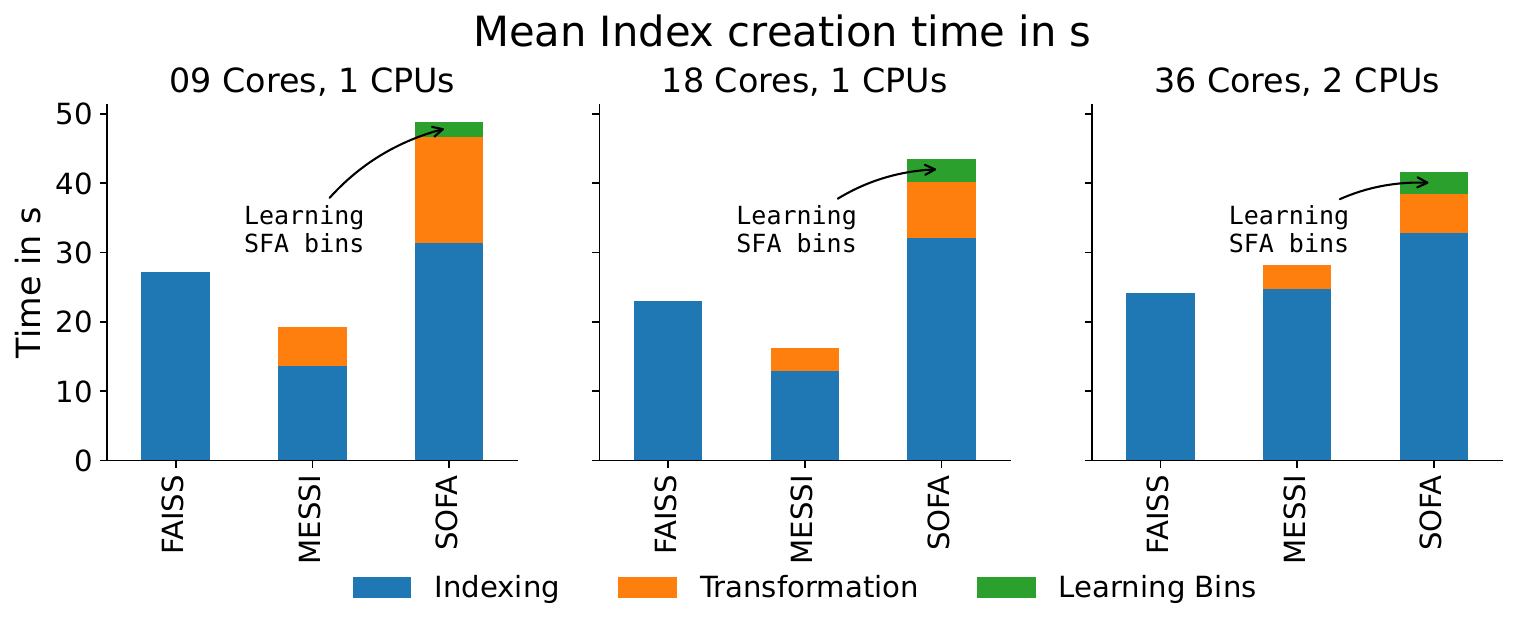}
	\caption{
	Comparison of mean indexing times using $9, 18$ or $36$ cores. MESSI is faster than SOFA. SFA involves some overhead for learning SFA bins (green) and the DFT (orange). Using more CPUs (18 to 36 cores) may increase runtime due to synchronization overhead.
	\label{fig:experiment_index_creation}
	}
\end{figure}

Figure~\ref{fig:experiment_index_creation} presents the mean runtime results for index creation across three systems: SOFA using SFA, MESSI using SAX for summarization, and FAISS (IndexFlatL2, CPU) using all 17 datasets. The measurements exclude I/O times for reading datasets from disk. 

On average, index creation times range from 15 seconds to 1 minute. MESSI is the fastest, averaging around 15 seconds, followed by FAISS. There is minimal improvement in index creation times when utilizing more cores. Notably, we observed an increase in index creation times when scaling from one CPU (18 cores) to two CPUs (36 cores), due to an increased overhead in synchronization.

The SFA method in SOFA incurs an overhead for learning the quantization from a $1\%$ sample of the data. This overhead is negligible compared to the time required for transformation and index creation. Overall, SFA has a higher transformation time than SAX, due to the use of the Fourier transform, which has a complexity of $\mathcal{O}(n \log n)$ in the DS length $n$, compared to $\mathcal{O}(n)$ for PAA. Index creation times are also higher for SOFA, potentially indicating more node splits. 

\begin{figure}[t]
    \centering  
	\includegraphics[width=1.00\columnwidth]{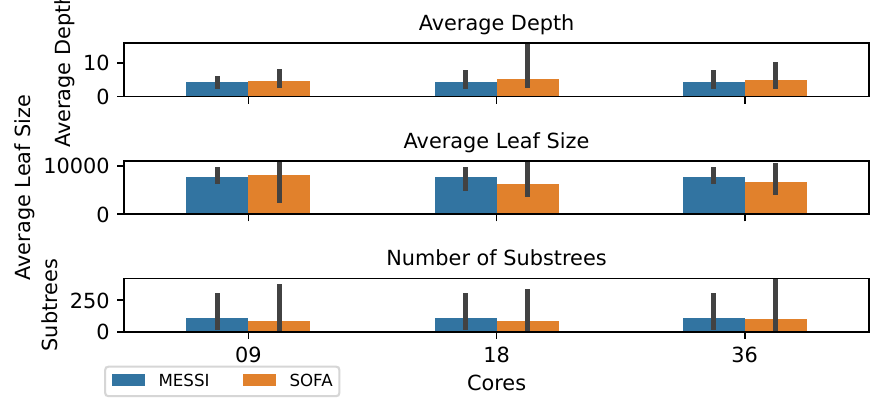}
	\caption{
	Index properties between MESSI and SOFA.
	\label{fig:experiment_index_prop}
	}
\end{figure}

Overall there is little difference in the structure of the indices over all $17$ datasets (Figure~\ref{fig:experiment_index_prop}). SOFA has a slightly higher average tree depth (top) and a smaller fill degree of the leafs (center). 
The fanout of the root node (bottom) is slightly lower for SOFA. 
For index construction, MESSI divides the data into chunks, processes each chunk independently, and then merges them into subtrees, which are subsequently combined into the full index. Therefore, the structure of the index may vary depending on the number of workers or cores used.

\subsection{Exact Similarity Search}\label{sec:experiments_exact_search}
\paragraph{1-NN Exact Search}
Table~\ref{tab:mean_query_times} shows the results for the exact 1-NN similarity search under ED. We report mean and median query times over all $17$ datasets. SOFA is the fastest method for all configurations, except for median on 9 cores, where FAISS is fastest. On average, SOFA is >10 times faster than UCR SUITE-P, 2-4 times faster than FAISS and 2-3 times faster than MESSI. All methods scale with the number of cores, except FAISS, which degenerates from 18 to 36 cores. With SOFA an exact 1-NN similarity query can be answered in $58$ ms in the median case on our hardware and with our datasets, faster than a blink of an eye.

\begin{table}
    \centering
\caption{Mean and Median 1-NN Query Times in ms for mixed workload on $17$ datasets. SOFA is fastest. \label{tab:mean_query_times}}
    \begin{NiceTabular}{c|c|cc}
    \toprule
     &  & median & mean \\
    Method & Cores &  &  \\
    \midrule
    \multirow[t]{3}{*}{FAISS IndexFlatL2~\cite{johnson2019billion}} 
     & 09 & 358 & \textbf{510} \\
     & 18 & 206 & 309 \\
     & 36 & 248 & 344 \\
    \cline{1-4}
    \multirow[t]{3}{*}{MESSI} 
     & 09 & 335 & 932 \\
     & 18 & 185 & 486 \\
     & 36 & 112 & 299 \\
    \cline{1-4}
    \multirow[t]{3}{*}{SOFA} 
     & 09 & \textbf{149} & 581 \\
     & 18 & \textbf{81} & \textbf{293} \\
     & 36 & \textbf{58} & \textbf{209} \\
    \cline{1-4}
    \multirow[t]{3}{*}{UCR SUITE-P~\cite{RakthanmanonCMBWZZK12}} 
     & 09 & 1448 & 1654 \\
     & 18 & 790 & 867 \\
     & 36 & 557 & 587 \\
    \cline{1-4}
    \end{NiceTabular}
\end{table}

\begin{figure}[t]
    \centering  
	\includegraphics[width=1.00\columnwidth]{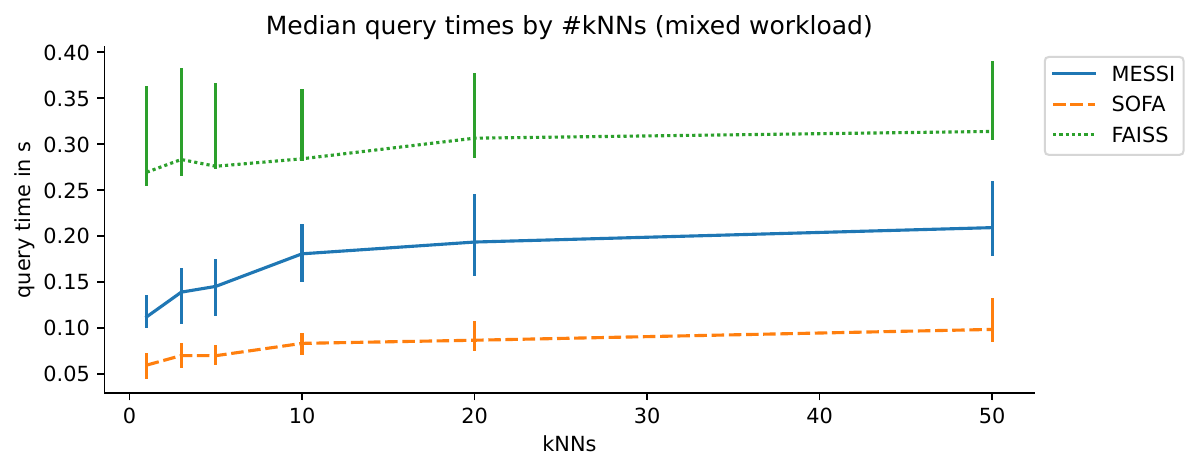}
	\caption{
	A comparison of median query times with an increasing number of nearest neighbors $k$. SOFA shows overall lowest query times. Notably, all methods scale efficiently with increasing $k$.
	\label{fig:scalability_knn}
	}
\end{figure}

\paragraph{k-NN Exact Search}
Table~\ref{tab:mean_query_times} presents the results for the exact $k$-NN similarity search using ED. We report the median query times across all $17$ datasets, utilizing $36$ cores. Among the methods, SOFA consistently demonstrated the fastest performance. For this experiment, we did not compute the full $k$-NN results for the UCR suite, as even for $1$-NN, its query times were already an order of magnitude larger compared to SOFA. Overall, SOFA exhibits the best scalability as the number of nearest neighbors increases~\ref{fig:scalability_knn}. Notably, all methods scale efficiently with increasing $k$.

\begin{table}
\caption{Median $k$-NN Query Times in ms for mixed workload on the $17$ datasets using $36$ cores. SOFA stays fastest. \label{tab:median_knn_query_times}}    
\begin{NiceTabular}{l|c|c|c|c|c|c}
    \toprule
    \textbf{Method} & \textbf{1-NN} & \textbf{3-NN} & \textbf{5-NN} & \textbf{10-NN} & \textbf{20-NN} & \textbf{50-NN} \\
    \midrule
    \text{UCR suite}    & 557 & - & - & - & - & - \\
    \text{FAISS}    & 248 & 283 & 276 & 284 & 307 & 314 \\
    \text{MESSI}    & 112 & 139 & 145 & 181 & 193 & 209 \\
    \text{SOFA}     & 58  & 70  & 70  & 83  & 87  & 98  \\
    \bottomrule
    \end{NiceTabular}
\end{table}

\begin{figure}[t]
    \centering  
	\includegraphics[width=1.00\columnwidth]{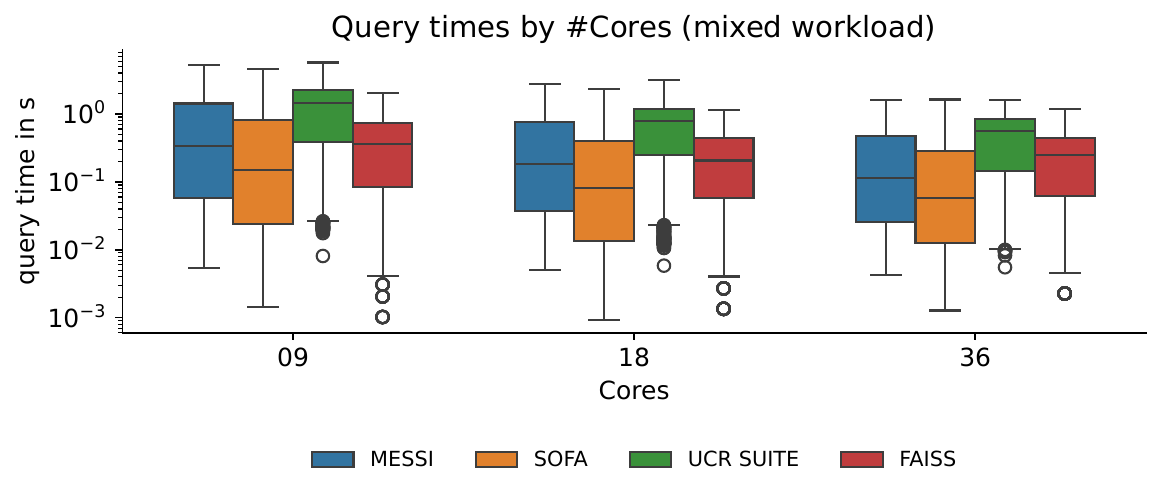}
	\caption{
	A comparison of 1-NN query times with an increasing number of cores on a logarithmic axis. SOFA shows overall lowest query times with most queries in the range of 100ms and some in the range of a few ms.
	\label{fig:scalability_cores_all}
	}
\end{figure}
\paragraph{Scalability in the number of cores} Figure~\ref{fig:scalability_cores_all} displays the query times for all datasets using box-plots, illustrating performance across an increasing number of cores. SOFA consistently shows the lowest median runtimes. However, both MESSI and SOFA exhibit high variance in query times across the different datasets. In contrast, the query times for FAISS and the UCR SUITE are more tightly clustered around their median values, indicating more consistent performance.

\begin{figure}[t]
    \centering  
	\includegraphics[width=1.00\columnwidth]{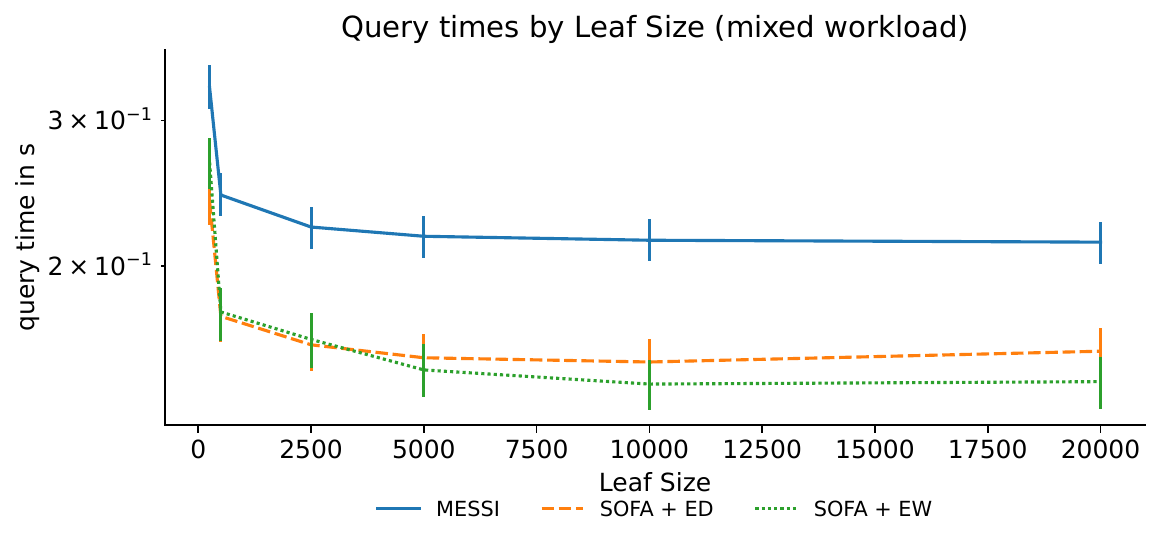}
	\caption{
	A comparison of 1-NN query times with an increasing leaf size. Query times decrease with increasing leaf node size and plateau around 10k data series.
	\label{fig:scalability_leaf_size}
	}
\end{figure}
\paragraph{Scalability in Leaf Size} Figure~\ref{fig:scalability_cores_all} displays the impact of increasing leaf sizes on the overall query times. With increasing leaf size, the query time decreases and plateaus around 10k data series. We used 20k as the default value for both MESSI and SOFA.

\begin{figure}[t]
    \centering  
	\includegraphics[width=1.00\columnwidth]{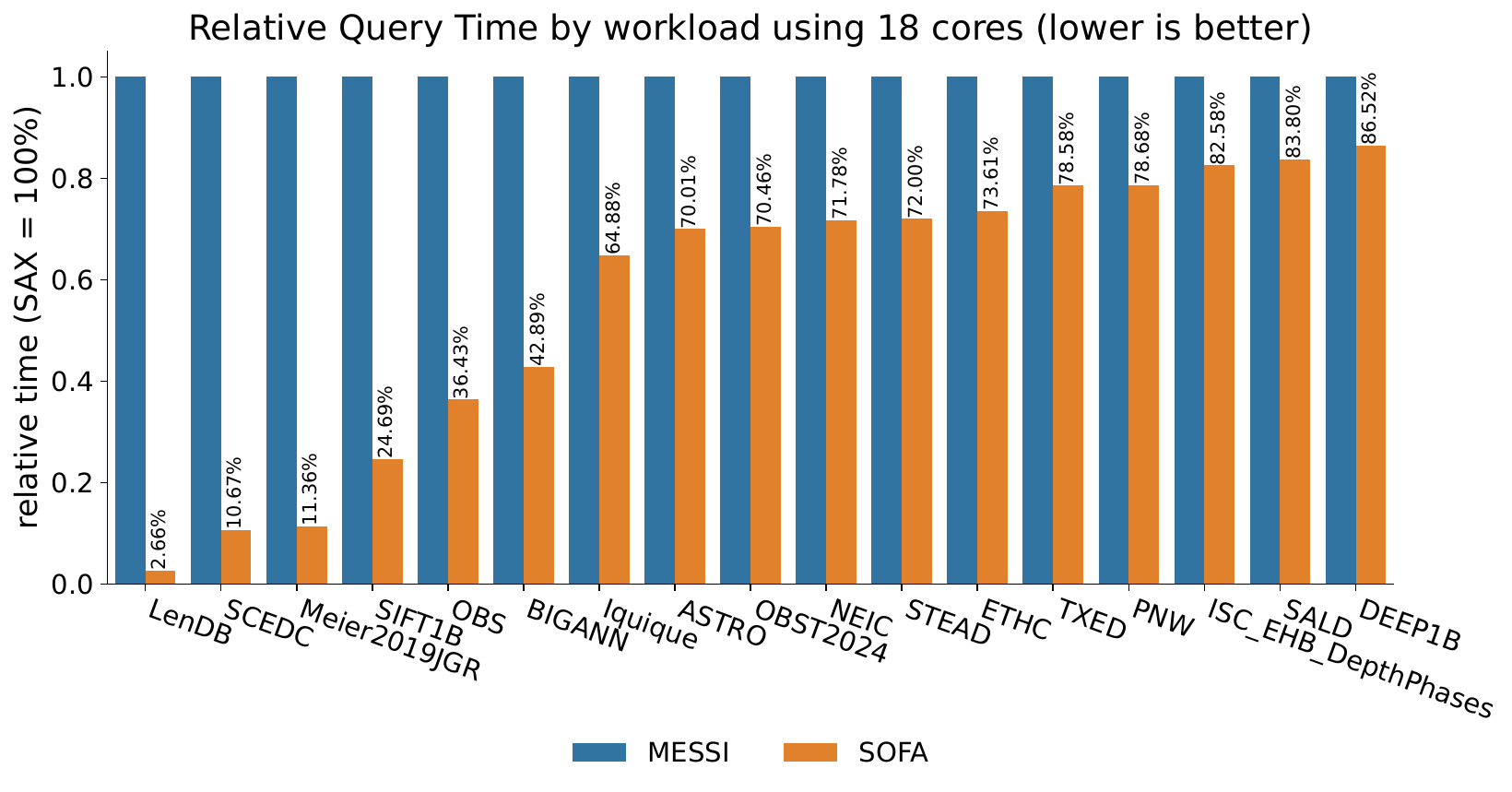}
	\caption{
	A comparison of relative query times for 1-NN using $18$ cores. The runtime improvement of SOFA over MESSI can be up to $38$ times faster (on LenDB).
	\label{fig:experiment_relative_query_time}
	}
\end{figure}

\paragraph{SOFA vs MESSI} To evaluate the impact of using SFA versus SAX, we conducted a comparative analysis between MESSI and SOFA. Figure~\ref{fig:experiment_relative_query_time} illustrates the relative improvement in average query times of SOFA compared to MESSI, with MESSI serving as the baseline ($100\%$). The results demonstrate that SOFA consistently outperforms its MESSI across all datasets, with some cases showing remarkable improvements of up to 38 times faster query processing (notably on the LenDB dataset).

Figure~\ref{fig:dataset_characteristics} highlights datasets where SOFA exhibits its most significant performance gains. In these cases, the SAX summarization technique employed by MESSI proves inadequate, as the mean-based segmentation results in oversimplified straight-line representations, particularly for datasets that do not follow a normal distribution. 
In contrast, SOFA's use of SFA allows for better adaptation to diverse data characteristics, resulting in substantially improved query times—up to $40$ times faster in some instances. 

This performance disparity underscores the superiority of SFA in scenarios with high-frequency DS, such as complex data patterns, and its ability to provide more accurate and efficient summarizations across a wide range of dataset types. The adaptive nature of SFA in SOFA proves particularly advantageous for datasets where traditional SAX-based approaches fall short, highlighting the importance of choosing appropriate summarization techniques for data series management systems. Our supporting website contains a comparison to FAISS~\cite{messi2}.

\subsection{Effect of Sampling on Query times}\label{sec:subsampling}

\begin{table}[t]
\centering
\caption{Performance of SOFA at different sampling rates on the $17$ datasets using $36$ cores.}
\label{tab:sofa_sampling}
\begin{NiceTabular}{l|c|c|c}
\textbf{Method} & \textbf{Sampling} & \textbf{Mean Time in ms} & \textbf{Median Time in ms} \\
\toprule
SOFA & 0.1\%  & 220 & 67 \\
     & 0.5\%  & 211 & 64 \\
     & 1\%    & 209 & 58 \\
     & 5\%    & 192 & 61 \\
     & 10\%   & 192 & 62 \\
     & 15\%   & 196 & 60 \\
     & 20\%   & 199 & 61 \\
\bottomrule
\end{NiceTabular}
\end{table}

SFA utilizes $1\%$ of the indexing samples to learn quantization bins. To explore the trade-offs associated with varying sampling rates, we conducted an experiment (Section~\ref{sec:messi_sfa}). As shown in Figure~\ref{tab:sofa_sampling}, the median query times stabilize at around a 1\% sampling rate, reaching 58 ms. However, the mean query times continue to improve up to a sampling rate of 5\%. Conversely, using less than 1\% of the data results in a slight increase in mean and median query times.

\subsection{When SOFA excels}

\begin{figure}[t]
    \centering      
    \includegraphics[width=0.6\columnwidth]{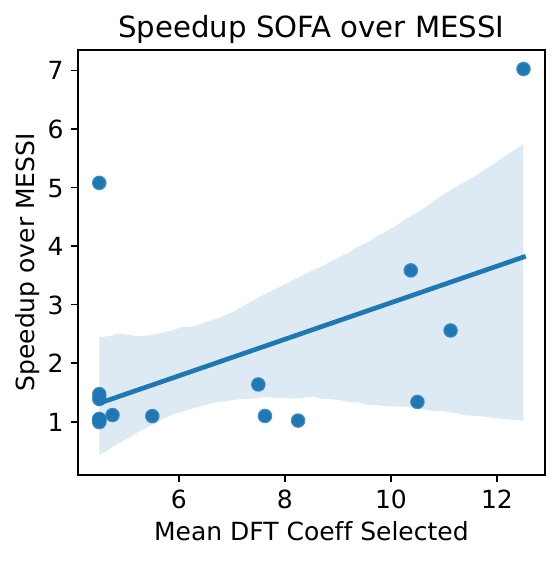}
	\caption{
	Comparison of the average index of the Fourier coefficients selected by SOFA against its speedup, relative to MESSI. Each point represents a single dataset. The results indicate a clear trend: higher selected frequencies are generally associated with greater speedup.
	\label{fig:experiment_variance}
	}
\end{figure}

SFA excels on datasets with rapid changes by selecting frequencies based on their largest variance (Section~\ref{sec:transform_and_learning}). To investigate the hypothesis that higher variance in frequency correlates with greater speedup, we examined the mean index of the Fourier coefficients selected by SOFA and its corresponding speedup over MESSI for each dataset. For instance, if SOFA selects eight Fourier coefficients with indices $[8, 9, 10, 11, 12, 13, 14, 15]$, their mean index would be $11.5$. The highest possible index that can be selected is $32$.

Our experiment (Figure~\ref{fig:experiment_variance}) reveals a clear trend, with a positive Pearson correlation of $0.51$. This suggests that Fourier coefficients corresponding to higher frequencies are associated with increased speedup—indicating that when SOFA prioritizes higher-frequency coefficients, it tends to outperform MESSI.
The datasets used in this experiment cover a wide range of frequency variances. Those characterized by low frequency variance include: Meier2019JGR, ASTRO, Iquique, NEIC, ETHZ, PNW, and SALD.

\subsection{Ablation Study}\label{sec:experiments_ablation}

We conducted an ablation study using the \emph{tightness of lower bound (TLB)} metric and two benchmark datasets. The TLB is defined as the LBD over the true distance~\cite{keogh2001dimensionality}. Higher is better, and the TLB reaches one if both are equal. A tighter (higher) TLB results in a better pruning of the search tree and thus a lower runtime. 

\begin{table}[ht]
\centering
\caption{Mean TLB on UCR datasets for increasing alphabet sizes.}\label{tab:ucr_tlb_results}
\begin{NiceTabular}{lrrrrrrr}
\toprule
Alphabet Size &       4   &       8   &       16  &       32  &       64  &       128 &       256 \\
Method      &           &           &           &           &           &           &           \\
\midrule
SFA ED +VAR &  0.65 &  0.74 &  0.78 &  0.80 &  0.81 &  0.81 &  0.81 \\
SFA EW +VAR &  0.62 &  0.72 &  0.77 &  0.80 &  0.81 &  0.82 &  0.82 \\
iSAX        &  0.48 &  0.60 &  0.66 &  0.71 &  0.73 &  0.75 &  0.76 \\
\bottomrule
\end{NiceTabular}
~\\~\\~\\
\caption{Mean TLB on SOFA datasets for increasing alphabet sizes.}\label{tab:sofa_tlb_results}
\begin{NiceTabular}{lrrrrrrr}
\toprule
Alphabet Size &       4   &       8   &       16  &       32  &       64  &       128 &       256 \\
Method      &           &           &           &           &           &           &           \\
\midrule
SFA ED +VAR &  0.41 &  0.49 &  0.54 &  0.57 &  0.58 &  0.60 &  0.61 \\
SFA EW +VAR &  0.34 &  0.47 &  0.54 &  0.59 &  0.61 &  0.63 &  0.64 \\
iSAX        &  0.37 &  0.45 &  0.45 &  0.52 &  0.54 &  0.55 &  0.55 \\
\bottomrule
\end{NiceTabular}
\label{tab:sofa_tlb}
\end{table}

\begin{enumerate}
    \item \textbf{UCR Archive}: We utilized the UCR dataset archive~\cite{dau2019ucr}, which consists of approximately 120 datasets spanning a wide range of applications. Each dataset includes a training and test set. The training set was used to learn the SFA representation, while the test set served as queries against the training data.
    \item \textbf{SOFA Datasets}: We also employed the 17 SOFA datasets used in this study (Table~\ref{tab:datasets}). Each dataset is divided into an indexing set, used to build the index, and a query set. The indexing set was used to learn the SFA representation, and the query set was used to perform queries against the indexing set.
\end{enumerate}

This allows us to evaluate TLB performance across diverse datasets and query scenarios.

\begin{figure}[t]
    \centering      
    \includegraphics[width=0.5\columnwidth]{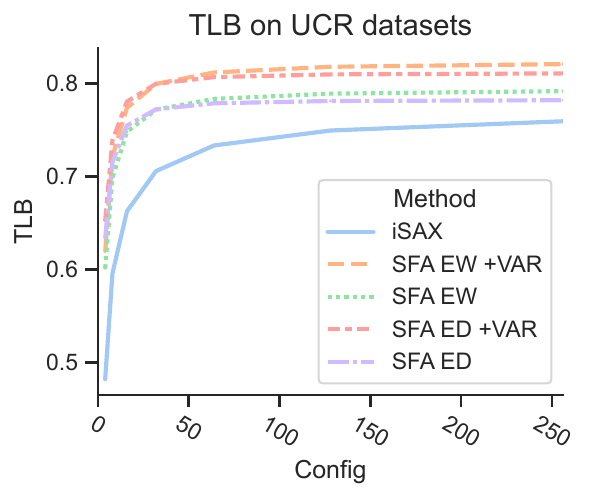}\includegraphics[width=0.5\columnwidth]{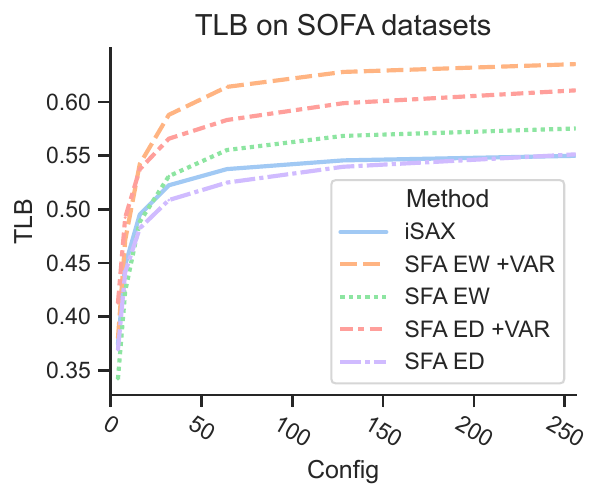}
	\caption{
	A comparison of the tightness of lower bound (TLB) on $120$ UCR datasets (left) and SOFA (right) datasets, for SFA variants against SAX for $l=16$. SOFA uses SFA with EW binning and a variance-based feature selection, the best performing method on both benchmarks.
	\label{fig:experiment_tlb}
	}
\end{figure}

For this study, we used SFA in combination with equi-width or equi-depth binning, with and or without using the variance for feature selection. For all summarizations we used length $l=16$. Figure~\ref{fig:experiment_tlb} shows the results for increasing alphabet sizes. Overall, SFA using equi-width binning for quantization and using variance for feature selection yields the highest TLB. Table~\ref{tab:ucr_tlb_results} and Table~\ref{tab:sofa_tlb_results} reveal that the improvement in TLB is particularly higher for smaller alphabet sizes, and is up to $17$ percentage points for alphabet size $4$ on the UCR datasets.

A high TLB enables the pruning of numerous ED calculations, which is one of the key reasons why SOFA outperforms MESSI in similarity search. It's important to note that distances to a query are not uniformly distributed, and even small improvements can result in substantial speedups. For instance, in the SCEDC dataset, which exhibits a 10x speedup (Figure~\ref{fig:experiment_relative_query_time}), the TLB difference is $24$ percentage points (pp) for an alphabet size of $4$. This translates into a significant pruning power difference~\cite{Chakrabarti2002} of 60 pp: specifically, we can prune $98\%$ of all data series at the first level of the tree, compared to $38\%$ with MESSI. Further pruning is achieved at subsequent levels as larger alphabet sizes are used.

\begin{figure}[t]
    \centering      
	\includegraphics[width=1.00\columnwidth]{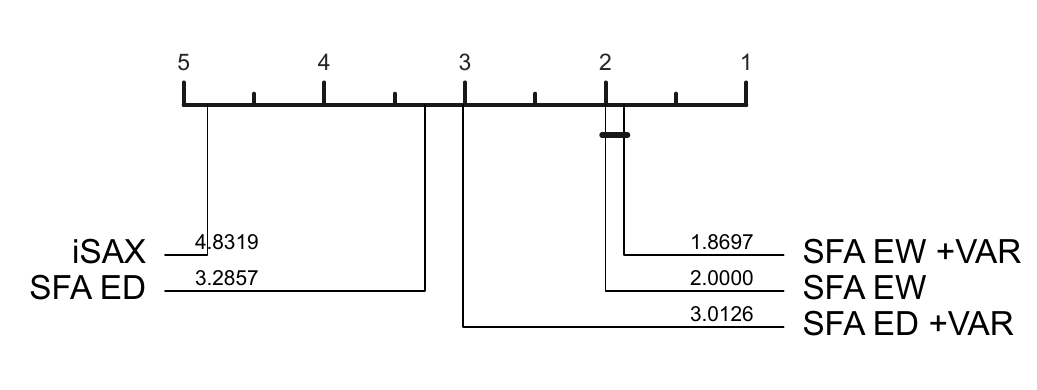}
    \includegraphics[width=1.00\columnwidth]{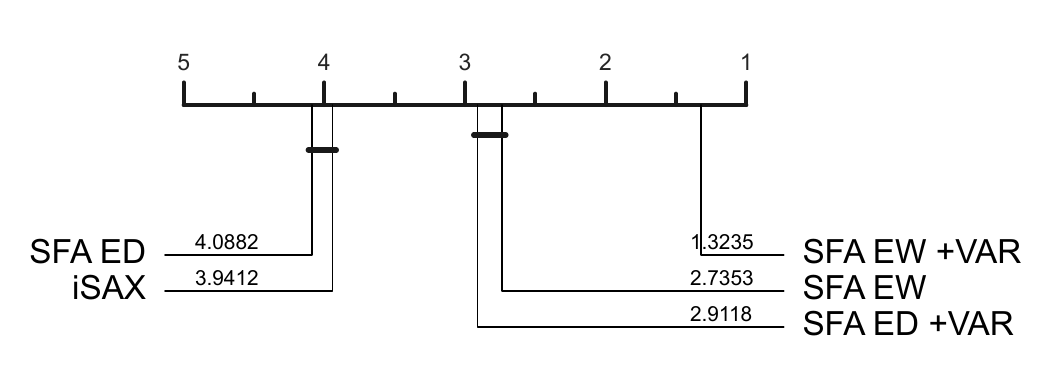}
    \vspace*{-0.8cm}
	\caption{
	Critical Difference plot on the average ranks of the TLB on 120 UCR datasets (top) and SOFA datasets (bottom) for $l=16$ and $\alpha=256$ (lower rank is better).
	\label{fig:experiment_tlb_ranks}
	}
\end{figure}

Finally, we use critical difference diagrams to compare mean ranks of approaches, with horizontal bars indicating statistically indistinguishable cliques (p-value 0.05, Wilcoxon-Holm post-hoc analysis). Figure~\ref{fig:experiment_tlb_ranks} shows that SFA EW+Var achieves significantly better ranks for both benchmarks, followed by ED and finally iSAX. Variance-based feature selection ensures optimal bin sizes. Equi-width outperforms equi-depth by avoiding small bins in z-normalized DS, which occur when using $256$ equi-depth bins in the -1 to +1 range.

\section{Conclusions}\label{sec:conclusion}

In this paper, we introduce SOFA, a fast and exact index for DS similarity search. We advocate the use of the Symbolic Fourier Approximation (SFA), a symbolization technique, which is learned from the actual data distribution in the frequency domain. 
SFA is tailored to high variance DS, where traditional approaches fail and lead to suboptimal performance. 
At the same time, SFA maintains its advantage across a variety of datasets.
Throughout our extensive experimental evaluation on a comprehensive benchmark comprising $17$ real use cases covering 1 billion DS and ~$1$TB, we show that SOFA is up to 38x faster than MESSI, and on average $2-10$ times faster than the SotA for exact similarity search. 

In our future work, we plan to study techniques for approximate similarity search using SFA, as well as parallelization opportunities through the use of GPUs.


\bibliographystyle{IEEEtranN}
\bibliography{messisfa}

\end{document}